\documentclass[fleqn,usenatbib]{mnras}

\usepackage{newtxtext,newtxmath}
\usepackage{xcolor}

\usepackage[T1]{fontenc}

\DeclareRobustCommand{\VAN}[3]{#2}
\let\VANthebibliography\thebibliography
\def\thebibliography{\DeclareRobustCommand{\VAN}[3]{##3}\VANthebibliography}

\usepackage{graphicx}	
\usepackage{amsmath}	
\usepackage{bbold}
\usepackage{comment}
\usepackage{bm}
\usepackage{mathtools}

\renewcommand{\vec}[1]{{\bm{#1}}}
\newcommand{\oper}[1]{{\bm{\mathsf{#1}}}}
\newcommand{\software}[1]{{\textsc{#1}}}

\title[Strong gravitational lensing in polarisation]{A self-consistent framework to study magnetic fields with strong gravitational lensing and polarised radio sources}

\author[Ndiritu et al.]{
S. Ndiritu,$^{1,2}$
S. Vegetti$^{1}$\thanks{E-mail: svegetti@mpa-garching.mpg.de},
D.~M. Powell$^{1}$ and  J.~P. McKean$^{2,3, 4}$
\\
$^{1}$Max Planck Institute for Astrophysics, Karl-Schwarzschild-Stra\ss{}e 1, 85748 Garching bei M\"unchen, Germany\\
$^{2}$Kapteyn Astronomical Institute, University of Groningen, Postbus 800, NL-9700 AV Groningen, The Netherlands\\
$^{3}$South African Radio Astronomy Observatory (SARAO), P.O. Box 443, Krugersdorp 1740, South Africa\\
$^{4}$Department of Physics, University of Pretoria, Lynnwood Road, Hatfield, Pretoria, 0083, South Africa
}

\date{Accepted XXX. Received YYY; in original form ZZZ}

\pubyear{2015}

\begin{document}
\label{firstpage}
\pagerange{\pageref{firstpage}--\pageref{lastpage}}
\maketitle

\begin{abstract}
We introduce a unified approach that, given a strong gravitationally lensed polarised source, self-consistently infers its complex surface brightness distribution and the lens galaxy mass-density profile, magnetic field and electron density from interferometric data. The method is fully Bayesian, pixellated and three-dimensional: the source light is reconstructed in each frequency channel on a Delaunay tessellation with a magnification-adaptive resolution. We tested this technique using simulated interferometric observations with a realistic model of the lens, for two different levels of source polarisation and two different lensing configurations. For all data sets, the presence of a Faraday rotating screen in the lens is supported by the data with strong statistical significance. In the region probed by the lensed images, we can recover the Rotation Measure and the parallel component of the magnetic field with an average error between 0.6 and 11 rad m$^{-2}$ and 0.3 and 3 nG, respectively. Given our choice of model, we find the electron density is the least well-constrained component due to a degeneracy with the magnetic field and disk inclination. 
The background source total intensity, polarisation fraction, and polarisation angle are inferred with an error between 4 and 10 per cent, 15 and 50 per cent, and 1 to 12 degrees, respectively. Our analysis shows that both the lensing configuration and the intrinsic model degeneracies play a role in the quality of the constraints that can be obtained. 
\end{abstract}




\section{Introduction}

Magnetic fields play an important role in the dynamics and star-formation processes of galaxies and, therefore, in the evolution of these objects \citep[e.g.][]{Hennebelle2019,Martin-Alvarez2020}. They are believed to grow through the dynamo effect \citep[e.g.][]{Kulsrud1999,Parker1992,Kulsrud2008,Martin-Alvarez2018,Rodrigues2019,Seta2020} from magnetic-seed fields, whose origin is still speculative. To this day, little is understood about their origin as well as their amplification \citep[e.g.][]{Rees2006}. 

To answer the many open questions regarding the formation and evolution of magnetic fields and how they affect various astrophysical processes, it is necessary to study their structural and statistical properties at early epochs. Owing to the sensitivity and angular resolution limitations of current observing facilities, direct measurements of the magnetic field structure in individual galaxies have so far mainly been limited to the local Universe \citep[e.g.][and references therein]{Lopez2023, Beck2013, Beck2015}.

Faraday rotation, the process by which the plane of polarisation for linearly polarised radiation propagating through a magneto-ionic medium is rotated \citep{Burn1966,Sokoloff1998,Beck2013}, is commonly used as an indirect tool for detecting and probing cosmic magnetic fields in distant galaxies \citep[e.g.][]{Kronberg1982,Welter1984,Oren1995,Kronberg2008,Bernet2008,Bernet2013,Farnes2017}. However, because of the unknown properties of the sources, and the unknown contribution of the Milky Way and the intergalactic medium (IGM), this approach results in a statistical measure of the magnetic field.

Strong gravitational lensing provides an independent channel to study magnetic fields in the individual distant lens galaxies \citep{Greenfield1985, Narasimha2008, Mao2017} and in the background sources \citep{Geach2023}. The basic idea is that strong gravitational lensing conserves the surface brightness, is achromatic and non-polarizing \citep{Dyer1992}, and, therefore, conserves the polarisation properties of the background object. This allows one to study magnetic fields in distant lensed sources with a signal-to-noise ratio and angular resolution that would not be otherwise possible at high redshifts \citep{Geach2023}. On the other hand, a non-homogeneous magnetised medium within the lens galaxy will induce a differential change in the observed polarisation properties of the lensed images. This propagation effect can be used to constrain the magnetic field and electron density in the lens galaxy, free from contamination by the Milky Way (at least for galaxy-scale lenses) and is independent of the intrinsic properties of the source \citep{Narasimha2008, Mao2017}. 

Due to the limited number of currently known strongly lensed polarised sources \citep[less than 10;][]{Greenfield1985,Biggs1999,Patnaik2001,Narasimha2008,Mao2017,Biggs2023}, this is a field that is still in its infancy, and that stands to significantly gain from the large number ($\sim10^5$) of strong gravitational lens systems that are expected to be discovered with the Square Kilometre Array \citep[SKA;][]{McKean2015} or the next generation Very Large Array (ngVLA). For example, extrapolating the fraction of polarised sources recently reported in the MeerKAT International GigaHertz Tiered Extragalactic (MIGHTEE) polarisation early-science data release \citep{Taylor2023}, one should expect $\sim2\times10^3$ lensed polarised sources detectable with the SKA at a surface brightness limit of 20 $\mu$Jy~beam$^{-1}$. 

Taking full advantage of future data will require lens modelling techniques that can infer the surface brightness distribution of the polarised source, as well as the mass distribution and magneto-ionic properties of the lens for a diverse population of lenses and sources. \citet{Burns2002} has proposed a method based on the misalignment between the polarisation vector and the source morphology (or any 2-d vector on the source plane) that is induced by the presence of a gravitational lens \citep[see also][]{Kronberg1991}. This approach only applies to background sources, such as radio jets and shocks, where these two quantities are expected to be intrinsically aligned on the source plane. Focusing on observations of lensed quasars,  \citet{Greenfield1985} and \citet{Mao2017} have decoupled the lens and the Faraday rotation analyses from each other and treated the lensed images of the same background unresolved object as they were, to a certain extent, independent. This procedure is unsuitable for resolved sources, as the multiple lensed images of the same source component cannot be trivially identified without a lens model.
Moreover, it assumes that the magnification does not change across the extent of the source and between the different lensed images, and it relies on \software{clean}ed maps for which the noise is correlated. In their analysis of a strongly lensed star-forming galaxy observed with the Atacama Large (sub)Milli-metre Array (ALMA), \citet{Geach2023} have first derived a model for the lens mass distribution using ancillary optical and millimetre data. This model was then used to lens forward a Gaussian model for the Stokes $\vec{Q}$ and $\vec{U}$ source surface brightness, assuming a constant polarisation angle. Due to the relatively high frequency of their observations (242 GHz), they could neglect the effect of Faraday rotation in the lens galaxy or along the line of sight to the background source.

In this paper, we introduce the first Bayesian gravitational lens modelling technique that, from the same data, self-consistently infers the lens galaxy mass distribution, magnetic field and electron density, and the polarised complex surface brightness distribution of the background source. The latter is modelled using a pixellated surface brightness distribution, requiring no strong assumptions on the morphological and polarisation properties of the source. Our technique is, therefore, suitable for modelling a large variety of lensed objects (e.g. galaxies and radio jets). The data are fitted in the visibility space where the noise is well-approximated by an uncorrelated Gaussian distribution. The method is three-dimensional (one frequency and two spatial dimensions) and fully forward; therefore, it automatically takes into account beam and bandwidth depolarisation.

The paper is organised as follows. In Section \ref{sec:method}, we provide a detailed description of our formalism, which we then test using simulated data that is generated according to Section \ref{sec:mock_data} and analysed using the modelling strategy outlined in Section \ref{sec:modelling}. We discuss the performance of the technique in Section \ref{sec:results} and its current limitations in Section \ref{sec:limitation}. We present our findings and a future outlook for the method in Section \ref{sec:summary}.

\section{Description of the method}
\label{sec:method} 

This section introduces our approach to analysing strong gravitational lensing observations with a polarised source and in the presence of a magnetised plasma in the lens galaxy. 

We assume that the foreground lensing galaxy is not a source of emission at the observed frequencies \citep[for example, 90 per cent of the foreground lenses found as part of the Cosmic Lens All-Sky Survey are radio-quiet;][]{Browne2003} and that the contribution to the Faraday rotation from the line of sight can be safely neglected. The latter assumption is justified by the fact that magnetic fields in the cosmic web are expected to be significantly smaller than those in galaxies \citep[e.g.][]{Carretti2022} and by the fact that the difference in the light-ray path lengths is much smaller than the path-length itself. As the angular separation of the multiple lensed images is small (a few arcsec on the plane of the observer), we also assume the effect of the Milky Way to be the same for all of the lensed images and, therefore, not significant. For simplicity, we limit our focus to the contribution of a regular large-scale magnetic field and defer the inclusion of a random component to a follow-up paper. We also ignore the effect of a possible field of axion-like particles on the observed polarisation angle \citep{Basu2021}. Under these conditions, we can treat any relative difference in the lensed image polarisation properties as arising from a single external Faraday rotating screen at the plane of the lens. At this stage, we assume the source will be constant throughout the observations and ignore the physical processes internal to this object. As discussed in the following, our formalism can be extended to include physical assumptions on the nature of the source in a self-consistent fashion. 

Our method builds upon \software{pronto}, the Bayesian grid-based software initially developed for optical data by \cite{Vegetti2009}. It was then further adapted for large interferometric data by \cite{Powell2021} and extended to the three-dimensional domain by  \cite{Rizzo2018}. Below, we describe in detail all new aspects of the method and refer the reader to the above papers as well as  \cite{Rybak2015a,Rybak2015b} and \cite{Ritondale2019} for further details on the original implementation and its subsequent developments. 

\subsection{The response operator}
\label{sec:response_op}

Given the electric field $\vec{e}= \left(e_{\rm R}(\vec{y}),e_{\rm L}(\vec{y})\right)^T$ of the background polarised source, the corresponding complex surface brightness distribution as a function of the position on the source plane $\vec{y}$ is given by the following coherency vector \citep{Hamaker1996, Smirnov2011a}
\begin{equation}
\vec{s}_{\rm{RL}} =
\begin{pmatrix}
\langle \vec{e}_{\rm R}\vec{e}^*_{\rm R}\rangle\\
\langle \vec{e}_{\rm R}\vec{e}^*_{\rm L}\rangle\\
\langle \vec{e}_{\rm L}\vec{e}^*_{\rm R}\rangle\\
\langle \vec{e}_{\rm L}\vec{e}^*_{\rm L}\rangle
\end{pmatrix}\,.
\end{equation}
Here, we have assumed that the instrument has circularly polarised feeds, which measure the right-handed (R) and left-handed (L) components of the incoming wave. Rather than working in the space of four complex components, we hereafter represent the source surface brightness distribution using the four real-valued Stokes parameters $\vec{I}$, $\vec{Q}$, $\vec{U}$, and $\vec{V}$:
\begin{equation}
\vec{s}\,  =\,  \begin{pmatrix}
\vec{I}\\
\vec{Q}\\
\vec{U}\\
\vec{V}
\end{pmatrix}\,  =\, 
\oper{F}_{\rm{RL}} \, \vec{s}_{\rm{RL}},
\label{eq:src_cvec}
\end{equation}
where $\oper{F}_{\rm{RL}}$ is the ``feed operator'' that forms the Stokes parameters from the appropriate linear combinations of raw RR, RL, LR, and LL correlations \citep{Hamaker1996}. Our method thus extends trivially to the case of orthogonal linearly-polarized (X and Y) feeds.

The gravitational effect of a strong gravitational lens galaxy between the source and the observer can be expressed in terms of the so-called lensing operator $\oper{L}$ \citep[see][]{Vegetti2009}, which is a function of the lens projected mass density distribution (see Section \ref{sec:mock_data} for a definition). It allows one to relate, via the lens equation, a position on the lens plane $\vec{x}$ to its corresponding position on the source plane $\vec{y}$ while taking into account the conservation of surface brightness. As described by \citet{Vegetti2009}, the latter is defined on a Delaunay tessellation, which is automatically adaptive with the lensing magnification. The lens plane is defined on a regular Cartesian grid of arbitrary resolution, though one should be careful to ensure Nyquist sampling. In the absence of a magnetised medium in the lens galaxy, the source light $\vec{s}_j$ and the observed visibility data $\vec{d}_j$ in each frequency channel $j$ are related to each other as follows:
\begin{equation}
    \vec{d}_j = \oper{D}_j~\oper{L}~\vec{s}_j + \vec{n}_j\,,
\label{eq:lin_sys}    
\end{equation}
where the NUFFT operator $\oper{D}_j$ includes, from left to right, a gridding operation, a Fast Fourier Transform, an apodisation correction, and a zero-padding/masking operation \citep[see][]{Powell2021}. Here, $\vec{n}_j$ is a vector encoding the observational noise, which is assumed to be Gaussian and uncorrelated between visibility points \citep{Wucknitz2002}. We assume the data $\vec{d}_j$ to be already calibrated and refer the reader to \citet{Smirnov2011a,Smirnov2011b} for a description of how to extend the response operator further to include calibration processes. From equations (\ref{eq:src_cvec}) and (\ref{eq:lin_sys}), it can be seen that the conservation of the polarisation properties (angle and fraction) of the background source by an intervening lens directly follows from the fact that lensing conserves surface brightness.

The main novelty of this paper is that we now include the effect of the lens galaxy magnetised interstellar medium (ISM), which acts as a single external Faraday screen and induces a differential rotation of the polarisation angle between the multiple lensed images of the same source \citep{Mao2017}. Taking into account this effect leads to a modified version of the forward model described by equation (\ref{eq:lin_sys}) as follows,
\begin{align}
    \vec{d}_j &= \oper{D}_j~\oper{S}_j~\oper{L} \, \vec{s}_j +\vec{n}_j \\
    &\equiv \oper{M}_j~\vec{s}_j +\vec{n}_j\,,
\label{eq:lin_sys_2} 
\end{align}
where we have introduced an external Faraday screen operator $\oper{S}_j$. In the Stokes basis, $\oper{S}_j$ takes the form
\begin{equation}
\oper{S}_j =
\begin{pmatrix}
\mathbb{1} & \mathbb{0} & \mathbb{0} &  \mathbb{0} \\
\mathbb{0} & \cos2\vec{\phi}_j & -\sin2\vec{\phi}_j & \mathbb{0}  \\
\mathbb{0} & \sin2\vec{\phi}_j & \cos2\vec{\phi}_j & \mathbb{0}  \\
\mathbb{0} & \mathbb{0} & \mathbb{0} & \mathbb{1}
\end{pmatrix}\,.
\label{eq:faraday_op}
\end{equation}
For a simple screen, the amount of rotation of the plane of polarisation is directly related to the Faraday depth in $\rm{rad \,m^{-2}}$ \citep[][]{Brentjens2005, Amaral2021}, so that
\begin{equation}
\vec{\phi}_j = \phi(\vec{x},\vec{B},\vec{n}_{\rm e}, \lambda_j^2) = \lambda_j^2 \times \mathrm{RM}(\vec{x},\vec{B},\vec{n}_{\rm e}).
\label{eq:faraday_angle}
\end{equation}
Equation \eqref{eq:faraday_angle} makes the wavelength-dependence ($\lambda_j$) of the polarisation rotation explicit; it is by modelling multiple frequency channels simultaneously that we can recover the properties of the Faraday screen. The Rotation Measure (RM) is
\begin{equation}
\mathrm{RM}(\vec{x},\vec{B},\vec{n}_{\rm e}) \equiv \frac{e^3}{2\pi m_{\rm e}^2c^4 \left(1+z_{\rm lens}\right)^{2}}\int{ n_{\rm e} (l, \vec{x}) B_{\rm LOS} (l,\vec{x})~dl}.
\label{eq:faraday_depth}
\end{equation}
Here, $n_{\rm e}(l,\vec{x})$ is the electron density, $B_{\rm LOS} (l,\vec{x})$\ is the line-of-sight component of the lens magnetic field, $dl$ is the path length, $z_{\rm{{lens}}}$ is the redshift of the lensing galaxy, $c$ is the speed of light, and $m_{\rm e}$ and $e$ are the rest mass and charge of the electron, respectively. 

\subsection{Bayesian inference}
\label{sec:bayes_inference}

In this section, we provide a summary of the Bayesian inference approach that allows us to simultaneously recover the source polarised surface brightness distribution $\vec{s}$, the lens mass distribution (see Section \ref{sec:mock_lens_mass} for a definition), the magnetic field and the electron density parameters (see Section \ref{sec:mock_lens_B} for a definition). For simplicity, we refer to the lens parameters (describing its mass distribution, magnetic field and electron density) collectively as $\vec{\eta}$. Our inference procedure is the same as the one introduced by \cite{Vegetti2009} and further developed by \cite{Rizzo2018} and \cite{Powell2021}. For completeness, we report the main key points here. In the following, we refer to $\vec{s}$ and $\vec{d}$ as the concatenation of the source and data vectors in the different frequency channels, that is, $\vec{s}=\{\vec{s}_{\rm 0}, ..,\vec{s}_{\rm nch}\}$ and $\vec{d}=\{\vec{d}_{\rm 0}, ..,\vec{d}_{\rm nch}\}$.

\begin{figure*}
\centering
\includegraphics[scale=0.5]{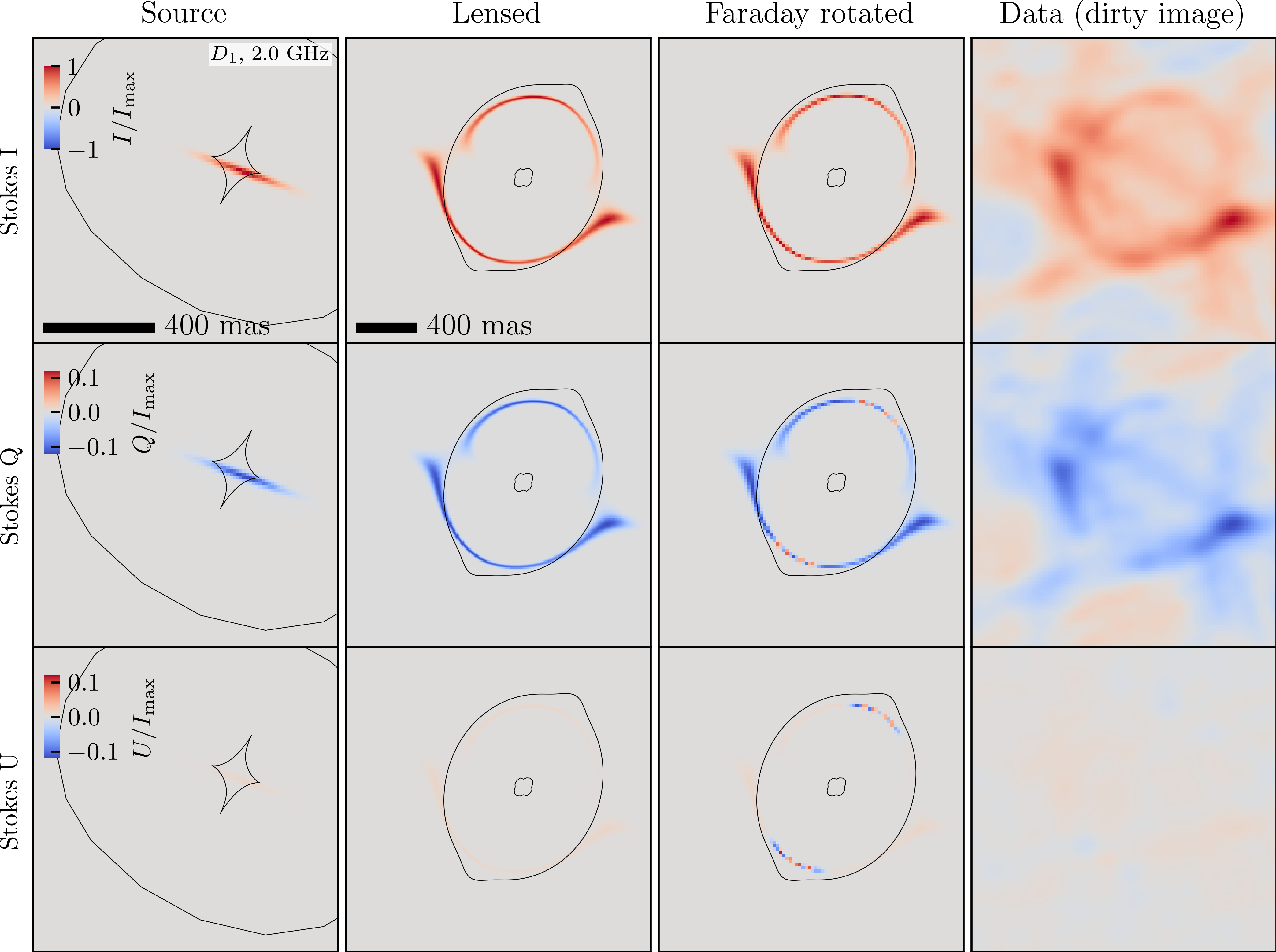}
\includegraphics[scale=0.5]{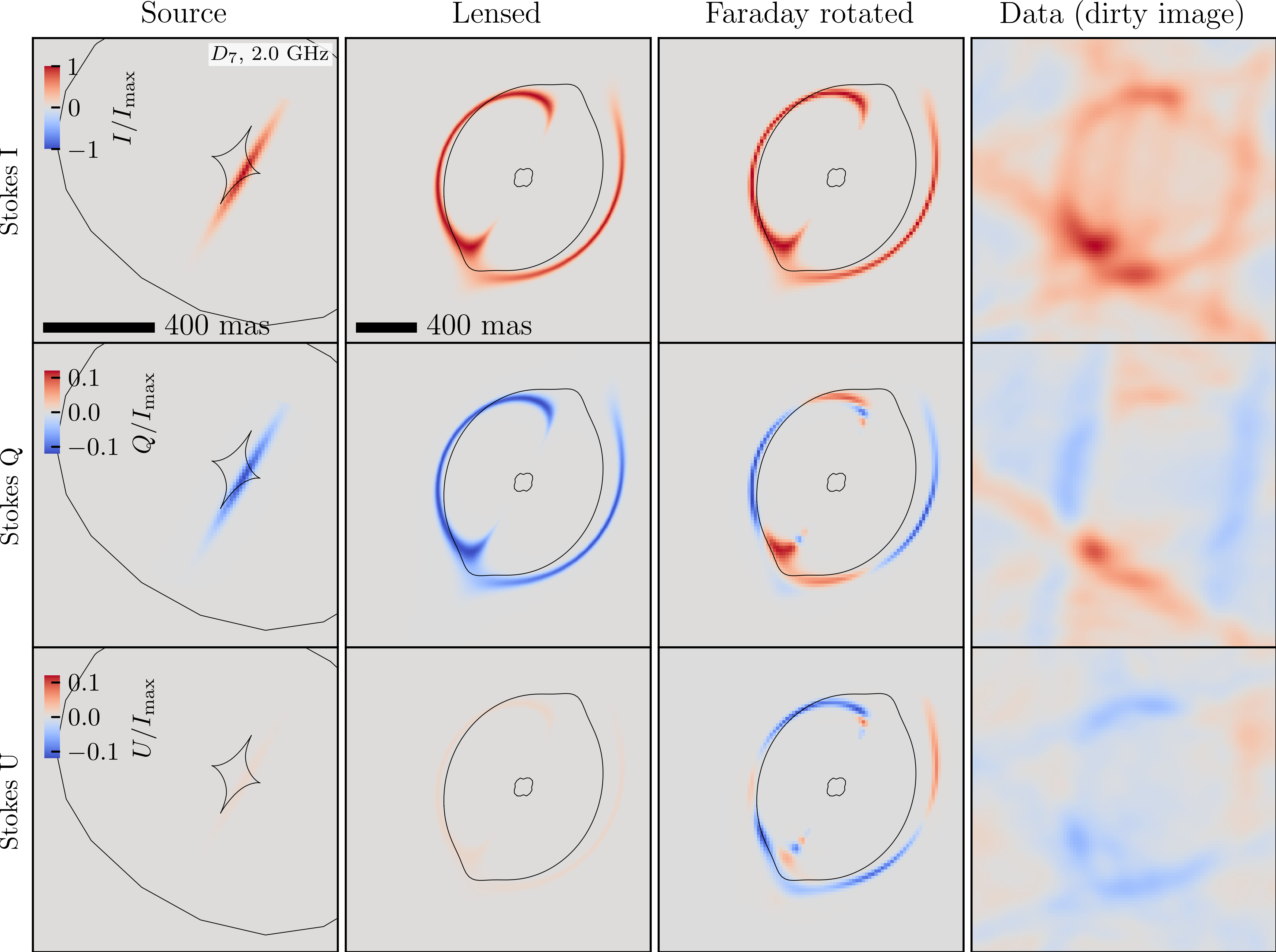}
\caption{Simulated data $D_{\rm 1}$ (source = $s_{\rm1}$, $P_f$ = 12 per cent, $i$ = 90 deg; left) and $D_\mathrm{7}$ (source = $s_{\rm2}$, $P_f$ = 12 per cent, $i$ = 45 deg; right) at the lowest frequency channel, as an example. From left to right: the input source, the lensed images, the Faraday-rotated lensed images, and the corresponding dirty images for the Stokes parameters $\vec{I}$ (top row), $\vec{Q}$ (middle row) and $\vec{U}$ (bottom row).}
\label{fig:mock_data_1}
\end{figure*}

\subsubsection{Source reconstruction}
\label{sec:src_inferece}

For a given choice of the lens parameters $\vec{\eta}$, we can infer the maximum posterior (MAP) source $\vec{s}_{\rm{MP}}$ by maximising the following probability:
\begin{equation}
P(\vec{s}|\vec{d}, \vec{\eta}, \lambda_{\rm s}, \oper{R}) =\frac{P(\vec{d}|\vec{s}, \vec{\eta})P(\vec{s}|\lambda_{\rm s}, \oper{R})}{P(\vec{d}|\lambda_{\rm s}, \vec{\eta},\oper{R})}\,.
\label{eq:src_posterior}
\end{equation}
Here, $P(\vec{d}|\vec{s},\vec{\eta})$ is the likelihood and $P(\vec{s}|\vec{\lambda}_{\rm s}, \oper{R})$ is a regularising prior of strength $\vec{\lambda}_{\rm s}$ (one per correlation) and order $\oper{R}$ on the source. Assuming the noise has a Gaussian distribution, we can express the likelihood function as: 
\begin{equation}
P(\vec{d}|\vec{s}, \vec{\eta}) = \frac{\exp [-E_{D}(\vec{d}|\vec{s},\vec{\eta})]}{\sqrt{\prod_{j}\det (2\pi \oper{C}_{j})}}\,,
\label{eq:likelihood}
\end{equation}
where $\oper{C}_{j}$ is the diagonal covariance matrix for the frequency channel $j$, and 
\begin{equation}
E_{D}(\vec{d}|\vec{s}, \vec{\eta}) = \frac{1}{2}\sum^{\rm{nch}}_{j}(\oper{M}_j~\vec{s}_{j} - \vec{d}_{j})^{T}\oper{C}^{-1}_{j}(\oper{M}\vec{s}_{j} -\vec{d}_{j})\,.
\label{eq:chi}
\end{equation}
The response operator $\oper{M}_j$ is given by equation (\ref{eq:lin_sys_2}). \cite{Vegetti2009} and \cite{Powell2021} have assumed a quadratic form for the source prior $P(\vec{s}|\vec{\lambda}_{\rm s}, \oper{R})$ with a zero mean, that is,
\begin{equation}
P(\vec{s}|\vec{\lambda}_{\rm{s}}, \oper{R}) = \frac{\exp [-\vec{\lambda}_{\rm s} E_{R}\left(\vec{s}| \oper{R}\right)]}{Z_{R}}\,,
\label{eq:src_prior}
\end{equation}
with
\begin{equation}
E_{R}\left(\vec{s}|\oper{R}\right) = \frac{1}{2}\sum^{\rm{nch}}_j\,\vec{s}_j^T \oper{H}_R\vec{s}_j\,,
\label{eq:src_prior_er}
\end{equation}
\begin{equation}
Z_{R}(\vec{\lambda}_{s}) = \int d^{N_s} \exp{\left[-\vec{\lambda}_{\rm s} E_{R}(s)\right]}\,,
\end{equation}
and
\begin{equation}\label{eq:08}
\oper{H}_{R} = \nabla \nabla E_{R} = \oper{R}^{T}\oper{R}\,.
\end{equation}
Under this assumption, maximising the posterior probability in equation (\ref{eq:src_posterior}) to obtain $\vec{s}_{\rm MP}$ reduces to solving the following system of linear equations for $\vec{s}_j$
\begin{equation}
\left[\oper{M}_j^{\rm T}\oper{C}_j^{-1}\oper{M}_j+\vec{\lambda}_{\rm s}\oper{H}_R\right]\vec{s}_j = \oper{M}^{\rm T}_j \oper{C}_j^{-1} \vec{d}_j\,.
\label{eq:src_linear}
\end{equation}
\cite{Rizzo2018} have further developed this approach to include an analytical and physically-motivated model $\vec{s}_{\rm hyp}$ as a hyper-prior to the pixellated source. In this case, equations (\ref{eq:src_prior_er}) and (\ref{eq:src_linear}) are respectively modified as follows:
\begin{equation}
E_{R}\left(\vec{s}| \oper{R}\right) = \sum^{\rm{nch}}_j\left[E_R(\vec{s}_{\mathrm{hyp},j} ) +\frac{1}{2}(\vec{s}_j - \vec{s}_{\mathrm{hyp},j})^{T} \oper{H}_R(\vec{s}_j-\vec{s}_{\mathrm{hyp},j)} \right]
\end{equation}
and 
\begin{equation}
\left[\oper{M}_j^{T}\oper{C}_j^{-1} \oper{M}_j + \vec{\lambda}_{\rm s} \oper{R}^T\oper{R}\right]\vec{s}_j = \oper{M}_j^{T}\oper{C}_j^{-1} \vec{d}_j + \vec{\lambda}_{\rm s}\oper{R}^T\oper{R}~\vec{s}_{\rm hyp \:j}\,.
\label{eq:src_linear_hyp}
\end{equation}
The hyper-parameters $\vec{\eta}_{\rm hyp}$ defining $\vec{s}_{\rm hyp}$ also become free (hyper) parameters of the model. Including a hyper-prior allows us to retain the freedom of a pixellated source while introducing a physical prior in a forward and self-consistent way. For example, $\vec{s}_{\rm hyp}$ could be related to the intrinsic properties of the source and its magnetic field, as well as include the effect of internal physical processes, such as time variations in the source polarisation properties. In this paper, we follow \citet{Powell2021} and assume $\vec{s}_{\rm hyp}=\vec{0}$. For generality, we include the source hyper-prior term in the relevant equations in the following section. 

\subsubsection{Non-linear parameters and model comparison}
\label{sec:non_linear}

We obtain the most probable parameters for the lens mass, magnetic field and electron density, and the source regularisation level by maximising the following posterior probability,
\begin{equation}
P(\vec{\eta}, \vec{\lambda}_{\rm s}, \vec{\eta}_{\rm hyp}|\vec{d}, \oper{R}) =\frac{P(\vec{d}|\vec{\eta}, \vec{\lambda}_{\rm s},\vec{\eta}_{\rm hyp}, \oper{R})P(\vec{\eta}, \vec{\lambda}_{\rm s}, \vec{\eta}_{\rm hyp})}{P(\vec{d}| \oper{R})}\,.
\label{eq:eta_posterior}
\end{equation}
If we assume a uniform prior for $\vec{\eta}$, $\log \vec{\lambda}_{\rm s}$ and $\vec{\eta}_{\rm hyp}$, maximising equation (\ref{eq:eta_posterior}) is equivalent to maximising the following expression \citep[see][for a derivation]{Rizzo2018},
\begin{multline}
    P(\vec{d}|\vec{\eta}, \vec{\lambda}_{\rm s}, \vec{\eta}_{\rm hyp}, \oper{R}) = -E(\vec{s}_{\rm MP}) - \frac{{\rm nch}}{2} \log \det \oper{H}_{E} + \frac{N_{\rm s}\rm{nch}}{2} \log \vec{\lambda}_{\rm s} \\ + \vec{\lambda}_{\rm s} E_{R}(\vec{s}_{\rm {hyp}}) +\frac{\rm{nch}}{2} \log \det \oper{H}_{R}  - \frac{N_{\rm{d}}\rm{nch}}{2} \log 2 \pi -\frac{1}{2}\sum^{\rm nch}_{j=1} \log \det \oper{C}_{j}\,,
\label{eq:max_non_linear}
\end{multline}
with $E = E_{D} +  \vec{\lambda}_{\rm s} E_{R}$. For each step in this non-linear optimisation, we calculate the corresponding maximum a posterior source $\vec{s}_{\rm MP}$ by solving the linear system (\ref{eq:src_linear}) or (\ref{eq:src_linear_hyp}), depending on the choice of hyper-prior.

As the magneto-ionic properties of distant galaxies are still poorly understood, it is important to consider and compare different models. Within our Bayesian inference approach, we can achieve this goal by comparing different assumptions for the magnetic field and electron density in terms of their Bayesian evidence, that is, the denominator in equation (\ref{eq:eta_posterior}), which we calculate using \software{MultiNest} by \citet{Feroz2019}. This also allows us to explore the multi-dimensional posterior distribution and identify potential degeneracy. Moreover, by comparing the Bayes factor of a model that includes the Faraday screen with one that does not, we can quantify the statistical significance with which the screen is detected in the first place.

\section{Simulated data}
\label{sec:mock_data}

\begin{table*}
	\centering
	\caption{Characteristics of the simulated lens systems for data sets D$_1$ to D$_8$ (given in the first column of each table). First Table: the parameters of the background source, that is, the redshift, total polarisation fraction, spectral index, size and position angle. Second Table: column 2 gives the redshift of the lens, while the other columns give the mass density parameters of the lens galaxy. Third and fourth Tables: the magnetic field and electron density parameters of the lens galaxy. All angles are defined as positive east of north.}
	\label{tab:mock_data}
	\begin{tabular}{l l l l l l l l l l l l l l l l}
  \hline
   \multicolumn{13}{c}{Input source parameters}\\
   \\
    Data set &z$_{\rm src}$ & $P_f$ & $P_\theta$ & $\alpha$ & Major axis & PA\\
    && [per cent] & [deg] & & [milli-arcsec] & [deg]\\
    \hline
     D$_{\rm 1}$ and D$_{\rm 5}$ 	& 1.34	& 12 	& 87	& $-$0.5 	& 10	& 70\\
     D$_{\rm 3}$ and D$_{\rm 7}$ 	& 1.34	& 12	& 87 	& $-$0.5 	& 15	& $-$30\\
     D$_{\rm 2}$ and D$_{\rm 6}$ 	& 1.34	& 3 	& 78	& $-$0.5	& 10	&  70\\
     D$_{\rm 4}$ and  D$_{\rm 8}$ 	& 1.34	& 3 	& 78 & $-$0.5 	& 15	& $-$30\\
\hline
    \multicolumn{13}{c}{Input lens parameters}\\
    \\
   Data set & z$_{\rm lens}$ & $\kappa_{\rm 0}$ & $\theta$ & $q$ & $r_{\rm c}$ & $\gamma$ & $\kappa_{\rm 0,d}$ & $q_{\rm d}$ & $R_{\rm d}$ & $\theta_{\rm d}$ &  $\Gamma$ & $\Gamma_{\mathrm{th}}$ \\
	   & & & [deg] &   &  [arcsec]   &     &    &     & [arcsec]  &  [deg]  & [arcsec]   &  [deg]\\
	\hline
     D$_{\rm 1}$ to  D$_{\rm 8}$ & 0.41 & 0.61 & $-$30 & 0.8 & 0.0001 & 2.0 & 0.77  & 0.20 & 0.19 & $-$30  & 0.03 & 23\\
    \hline
   \multicolumn{13}{c}{Input magnetic field parameters}\\
   \\
  Data set & $B_{\rm 0}$ & $h$ & $r_{\rm 0}$ & $\phi_{\rm 0}$ &  $p$ & $\chi_{\rm 0}$ & $i$ & $\theta$\\
  & [$\mu$G] & [kpc] & [kpc] & [deg] & [deg] & [deg] & [deg] & [deg]\\
 
 \hline
     D$_{\rm 1}$ to  D$_{\rm 4}$ & 11 & 1 & 10 & 80 & 10 & 45 & 90 & $-$30\\
     D$_{\rm 5}$ to  D$_{\rm 8}$ & 11  & 1 & 10 & 80 & 10 & 45  & 45 & $-$30\\
    \hline 
   \multicolumn{13}{c}{Input electron density parameters}\\
   \\
   Data set & $n_{\rm 0}$ & $h_{\rm ne}$ & $r_{\rm ne}$ & $i$ & $\theta$\\
    & [cm$^{-3}$] & [kpc] & [kpc] & [deg] & [deg]\\
	\hline
    D$_{\rm 1}$ to  D$_{\rm 4}$  & 0.03 & 1 & 10 & 90 & $-$30\\
    D$_{\rm 5}$ to  D$_{\rm 8}$  & 0.03 & 1 & 10 & 45 & $-$30\\
   \hline
    \end{tabular}
\end{table*}

This section describes the process for creating eight simulated interferometric observations.
Fig.~\ref{fig:mock_data_1} shows the steps in the data simulation pipeline, going from the input surface brightness distribution of the polarised source to the (not deconvolved) images produced by an interferometer for two different data sets. In Section \ref{sec:mock_src}, we introduce our models for the background source, which we then lens forward via the mass-density model described in Section \ref{sec:mock_lens_mass}. The lensed images are then Faraday rotated according to the magnetic field and electron density models provided in Section \ref{sec:mock_lens_B}. Table \ref{tab:mock_data} lists our choice of input lens and source parameters. Details on the simulated observations are given in Section \ref{sec:mock_obs}. 

\subsection{Source model}
\label{sec:mock_src}

The input source model has a Gaussian surface brightness profile with a total flux-density $S_\nu$ that varies with frequency $\nu$ via the power-law $S_{\rm \nu}\propto\nu^\alpha$, where the spectral index is set to $\alpha= -0.5$. This choice of spectral index is consistent with a flat-spectrum radio source but is not expected to strongly influence our results since it mainly ensures a close to uniform signal-to-noise ratio as a function of frequency. Our goal is to constrain the RM in different scenarios. Therefore, we consider two different source polarisation fractions ($P_f$) and two different source polarisation angles ($P_\theta$), which are defined as, 
\begin{equation}
    P_{\rm f}\left(x, y,\lambda\right) = \frac{\sqrt{\vec{Q}^2+\vec{U}^2+\vec{V}^2}}{\vec{I}} 
\end{equation}
and
\begin{equation}
    P_\theta\left(x, y,\lambda\right) = \frac{1}{2} \arctan{\frac{\vec{U}}{\vec{Q}}}\,,
\end{equation}
respectively. For the former, we consider a low (3 per cent) and a high (12 per cent) polarisation fraction case, and for the latter, we use a polarisation angle of 87 and 78 deg. We also consider two different lensing configurations derived from the same input mass model. These are obtained from two different source sizes and orientations of the major axis relative to the lens mass density distribution. For the rest of this paper, we refer to the source that has the smaller major-axis (10 mas) with a position angle (PA) that is misaligned relative to the lens (PA = 70 deg) as $s_{\rm1}$, while $s_{\rm2}$ has a larger major-axis (15 mas) and is aligned with the mass distribution (PA = $-$30 deg) to maximise the radial structure in the lensed images (see Fig.~\ref{fig:mock_data_1}). Our choice of source properties and lens parameters leads to a larger fraction of $s_{\rm1}$ than $s_{\rm2}$ being quadruply imaged. In all cases, we assume a linearly polarised source and set Stokes $\vec{V}$ to zero. For each Stokes parameter, we generate a total of 13 frequency channels between 2 and 4 GHz.

\subsection{Lens model}
\label{sec:mock_lens}

\begin{figure*}
\centering
\includegraphics[scale=0.6]{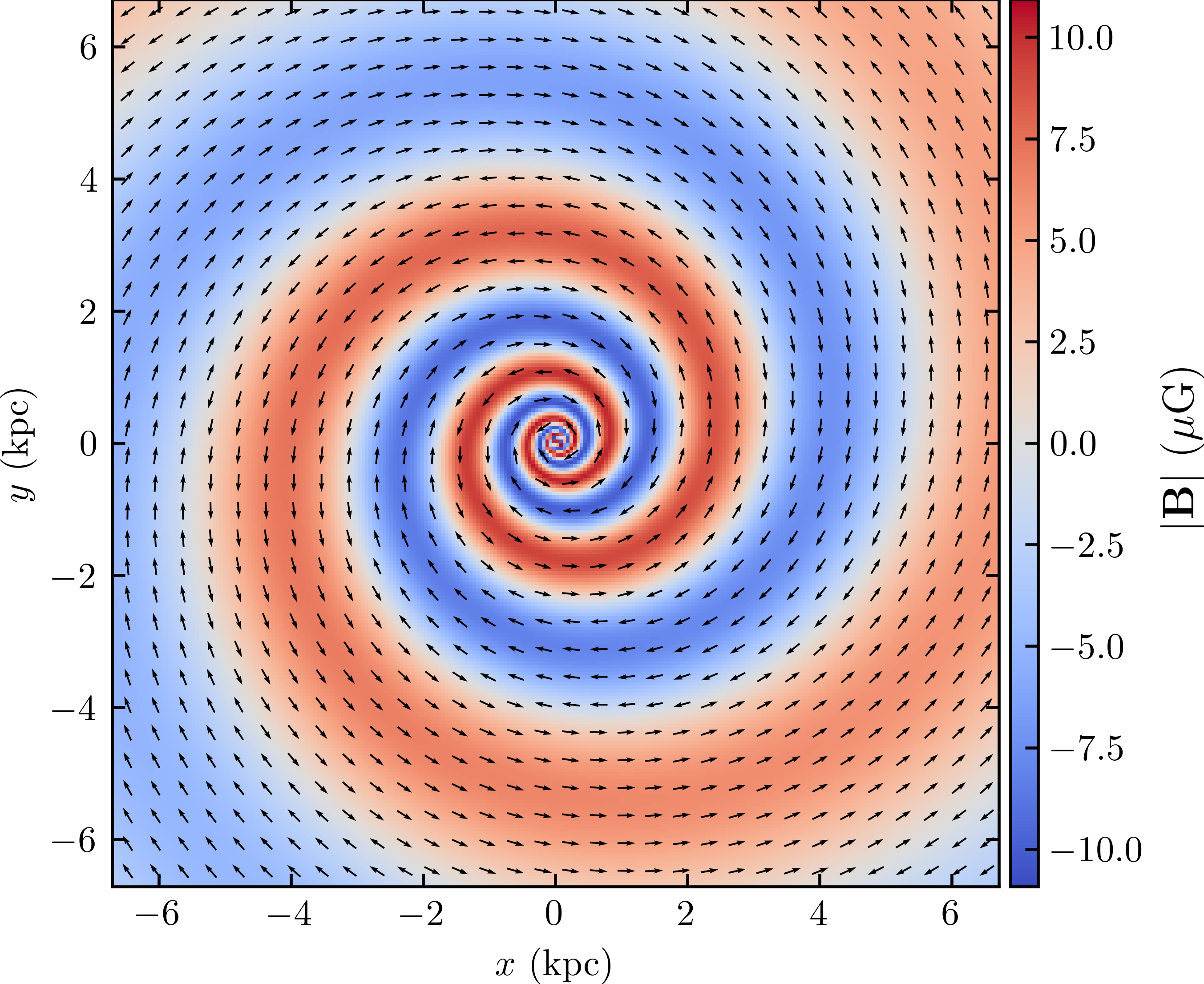}
\includegraphics[scale=0.6]{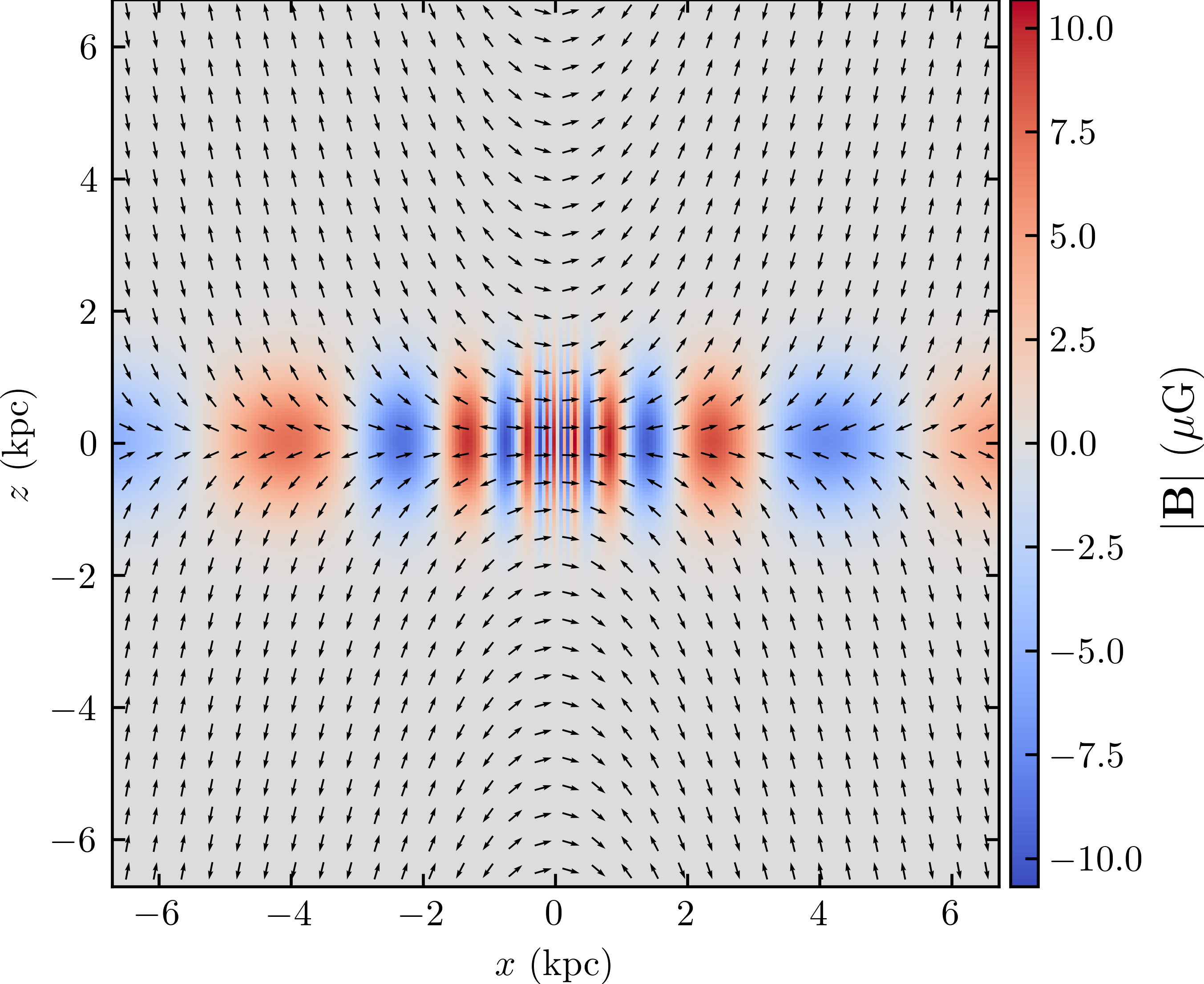}
\caption{Magnetic field model projections on the xy (left) and xz (right) planes. The colour scale is the amplitude $B_{\rm m}$ (eq. \ref{eq:BASS}), and the arrows indicate the direction of the field on the plane.}
\label{fig:bfield}
\end{figure*}

\subsubsection{Mass model}
\label{sec:mock_lens_mass}

We consider a two-component model for the mass density distribution of the lens galaxy, made of a single elliptical power law plus an edge-on disk for all of the simulated data. The lensing convergence for the power-law term is defined as,
\begin{equation}
\kappa_{\rm PL}\left(\rho\right) = \frac{\kappa_{\rm 0}\left(2-\frac{\gamma}{2}\right) q^{-1/2}}{2\left(r_{\rm c}^2+\rho^2\right)^{\left( \gamma-1\right)/2}}\,,
\label{eq:kappa_pl}
\end{equation}
where $\kappa_{\rm 0}$ is the amplitude, $\gamma$ is the 3D slope, $q$ is the projected axial ratio and $r_{\rm c}$ is the core radius. Together with the position angle $\theta$ and the position coordinates $x$ and $y$, they represent the geometrical parameters of the lens mass model. Following \citet{Keeton2011}, we approximate the convergence of an elliptical edge-on disk component as the sum of Kuzmin disks, such that,
\begin{equation}
\kappa_{\rm disk}\left(\rho\right) = \frac{\kappa_{\rm 0,d}}{q_{\rm d}}\sum_i^{11}\kappa_i\left(R_{\rm d} s_i\right)^3 \left(\left(R_{\rm 0} s_i\right)^2+\rho^2\right)^{3/2}\,,
\label{eq:kappa_disk}
\end{equation}
with the coefficients $s_i$ and $\kappa_i$ re-fitted using those parameters of the lens system B0712$+$472, as determined by \citet{Hsueh2017}, which is known to have an edge-on disk that crosses the lensed images. In the expression above, $\kappa_{\rm 0,d}$ is the central convergence, $R_{\rm d}$ is the scale length, and $q_{\rm d}$ is the axis-ratio. We also include an external shear component defined by a strength $\Gamma$ and a position angle $\Gamma_\theta$. Our choice of input parameters for the lensing mass distribution is based on modelling the real gravitational lens system B0712$+$472 \citep{Hsueh2017}.

For consistency, we use the same lens mass model for all of the data sets investigated here since \citet{Powell2021} have already demonstrated that our lens modelling technique can robustly recover the lens model parameters. Due to the different source surface brightness distributions, this same mass model will produce different configurations of the lensed emission (see Fig.~\ref{fig:mock_data_1}).

\subsubsection{Magnetic field and electron density model}
\label{sec:mock_lens_B}

We adopt a bi-symmetric azimuthal magnetic-field model with a vertical extension for all simulated data sets to account for the disk component (see Fig.~\ref{fig:bfield}). Our expression is a slight modification from that presented by \citet{Stepanov2008}, and in cylindrical coordinates, is defined as,
\begin{equation}
\begin{array}{@{}l@{}}
B_{\rm r}    = B_{\rm m}\sin{p}\cos{\chi}\\
B_{\rm \phi} = B_{\rm m}\cos{p}\cos{\chi}\\
B_{\rm z}    = B_{\rm m}\sin{\chi}
\end{array}
\end{equation}
where $p$ and  $\chi$ are the pitch and tilt angles, respectively. $B_{\rm m}$ and $\chi$ are respectively given by,
\begin{equation}
B_{\rm m} = B_{\rm 0}\exp{\left(-\frac{r}{r_{\rm 0}}\right)}\exp\left\{-\left(\frac{z}{h}\right)^2\right\}\cos{\left(\frac{\ln{r}}{\tan{p}}-\phi+\phi_{\rm 0}\right)}
\label{eq:BASS}
\end{equation}
and 
\begin{equation}
\chi = \chi_{\rm 0} \tanh{\frac{z}{2h}}\tanh{\frac{3r}{r_{\rm 0}}}\,.
\end{equation}
Following \citet{Stepanov2008}, we consider a Gaussian profile for the electron density, such that,
\begin{equation}
n_{\rm e}(r,z)= n_{\rm 0} \exp\left\{-\left(\frac{z}{h_{\rm ne}}\right)^2\right\}\exp\left\{-\left(\frac{r}{r_{\rm ne}}\right)^2\right\} .
\label{eq:ne_disk}
\end{equation}
The disk component in both $\vec{B}$ and $\vec{n_{\rm e}}$ have a position angle that matches the disk component of the lens mass density distribution. We set the inclination angle to 90 deg for half of the data sets and 45 deg for the other half. This leads to a rotating screen that is stronger (peak RM$_{\rm max} \sim$ 300 rad~m$^{-2}$) but is also more localised (i.e., it affects a smaller portion of the lensed images, and it fluctuates on smaller spatial scales) in the first case, and weaker (peak RM$_{\rm max} \sim$ 120 rad m$^{-2}$), but more extended and slowly changing for the second one. For the disk component with a 45 deg inclination angle, we do not re-align the mass model with $\vec{B}$ and $\vec{n_{\rm e}}$ because this would change the lensing configuration, preventing a direct comparison in terms of the Faraday screen. In all cases, we choose a value of $B_{0}$, consistent with the mean coherent magnetic field strength inferred by \citet{Mao2017} for the gravitational lens system B1152+199. All other parameter values are taken from \citet{Stepanov2008}. 

It is important to note that whether these models are a good representation of distant lens galaxies is currently unknown. In the future, one can explicitly test these assumptions from observations of the many strong gravitational lens systems expected to be discovered. This paper aims to test our modelling approach in a reasonably realistic scenario for which the above models are sufficient.

\subsection{Observations}
\label{sec:mock_obs}

The final part of our simulation pipeline adds the response of the interferometer. This is done to simulate an observed data set that is representative of the angular resolution and sensitivity of next-generation instruments, which are currently being constructed or are in development. However, the actual array configuration and receiver deployment have still to be finalised for the SKA-MID and the ngVLA. From the model surface brightness distributions presented in Fig.~\ref{fig:mock_data_1}, we see that the inclusion of the disk component results in large-scale changes to the Stokes $\vec{Q}$ and $\vec{U}$ emission that are on the order of 100 to 200 mas in size; we use this to inform the minimum angular resolution needed to identify these variations in the observed data, which can be achieved through a combination of the baseline length and the choice of observing frequency. 

We generated visibility data with an effective baseline length of about 150 km; this is consistent with the maximum baseline length of the Array Assembly 4 (AA4) of the SKA, which is expected to be completed by 2030 and should be an order of magnitude more sensitive than the VLA. We note that observing at a higher frequency can compensate for a shorter maximum baseline length. Our frequency choice is based on having a sufficient fractional bandwidth ($>50$ per cent) from which the Faraday rotation of the lensed emission can be reliably detected while providing a high enough angular resolution (150 mas) so that beam de-polarization does not overly affect our analysis. For this, we choose observing frequencies between 2 and 4 GHz. However, once the final array configuration and receiver deployment of the SKA-MID are fixed, we will perform more realistic simulations to determine the appropriate observational setup in terms of observing frequency, bandwidth and depth. Here, we aim to validate our method while providing some indication of the accuracy and precision with which the magneto-ionic medium within intermediate redshift galaxies can be recovered.

We create the data sets with the Common Astronomy Software Applications (CASA), using a template that is based on observations taken with the VLA, except that the maximum baseline length was increased from 36 to 150 km. This interferometer has 27 antennas that provide a good snapshot {\it uv}-coverage and produces 351 baselines per frequency and time sampling. As discussed above, we used 13 frequency channels to describe the 2 GHz of bandwidth between 2 and 4 GHz. The visibility integration time is 2 seconds, and the total integration time is 6 minutes. Overall, this produced a data set with around 820\,000 visibilities per correlations (RR, LL, RL and LR), which is sufficiently small for testing our method.

\section{Modelling strategy}
\label{sec:modelling}

We model each data set using the pixellated and regularised method described in Section \ref{sec:method}. The free parameters of the model are those describing the lens mass density distribution, magnetic field and electron density, and the source surface brightness profile. At every stage of the modelling procedure, we keep the core radius of the mass density fixed, and we force the magnetic field and electron density to be aligned and centred with each other. As the normalisations $B_{\rm 0}$ and $n_{\rm 0}$ (see equations \ref{eq:BASS} and \ref{eq:ne_disk}) are degenerate with each other, we refactor them into a single free parameter  $A_{\rm 0}=B_{\rm 0} \times n_{\rm 0}$. 
 
One also has to infer the regularisation level $\vec{\lambda}_{\rm s} = \{\lambda_{\rm s,I}, \lambda_{\rm s,Q}, \lambda_{\rm s,U} \lambda_{\rm s,V}\}$ for a given form of regularisation $\oper{R}$. Given the significant difference in signal-to-noise ratio between the Stokes parameters, we found that leaving the regularisation constants to change independently from one another is necessary to obtain a satisfactory source reconstruction for each polarisation component. 

From equation (\ref{eq:faraday_op}),  it can be seen that the Stokes $\vec{I}$ surface brightness, which provides the constraints to the lensing potential, is unaffected by the presence of an ionised medium in the lens. We note that interstellar scattering or free-free absorption from free electrons can affect the Stokes $\vec{I}$ emission, but these are the focus of a companion paper (Kaung et al., in prep.) and are not considered further here. As a result, we expect the parameters describing the lens mass-density distribution and its magneto-ionic properties to be largely independent of each other. Hence, we model the data in four stages following the Bayesian inference approach described in Section \ref{sec:bayes_inference}. (i) First, we neglect the contribution of the lens as a Faraday rotating screen and infer the MAP lens mass density distribution, source and source regularisation level $\lambda_{\rm s,I}$ from the Stokes $\vec{I}$. (ii) We then keep the lens mass parameters fixed to the results obtained from step (i), include the effect of the Faraday-rotating screen and obtain the MAP of the lens magnetic field, electron density, the source and its regularisation levels by modelling all Stokes parameters, as described in Section \ref{sec:bayes_inference}. (iii) We use \software{MultiNest} to quantify the statistical errors on all of the parameters assuming a uniform prior for $\vec{\eta}$ and $\log{\vec{\lambda}_s}$. (iv) By comparing the Bayesian evidence of those models with and without the contribution of the Faraday screen, we obtain a measure of the statistical significance of the magnetised plasma within the foreground lens galaxy.

\section{Results}
\label{sec:results}

\begin{figure*}
\centering
\includegraphics[angle=90,width=0.9\textwidth]{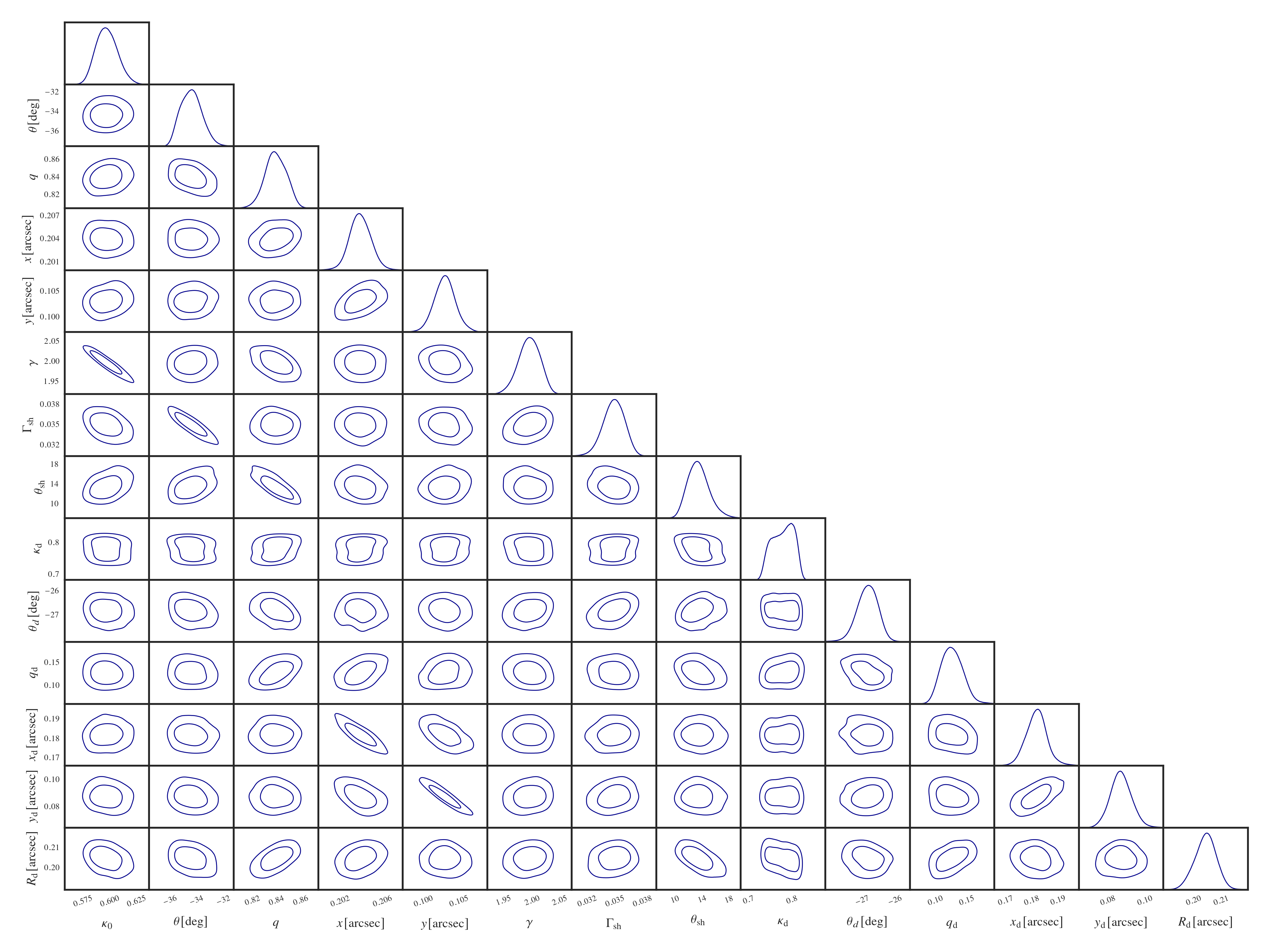}
\caption{The posterior distribution for all parameters of the lens mass density model for data set $D_{\rm 6}$ (source = $s_{\rm1}$, $P_f$ = 3 per cent, $i = 45$~deg).}
\label{fig:mn_lens}
\end{figure*}

In this section, we present our results for the simultaneous modelling of the mass distribution and magneto-ionic medium of the lens, as well as the properties of the recovered background source. We quantify the quality of our reconstructions of each model parameter $x$ in terms of the root-mean-square residuals (i.e. $\Delta x$) and the root-mean-square residuals normalised to the peak input value (i.e. $\sigma_x$). Both quantities are averaged over the region where the lensed images are at least 20 per cent of the peak surface brightness. Our results are summarised in Figs. \ref{fig:mn_lens} to \ref{fig:SLs_comparison} and in Tables \ref{tab:multinest_mass} and \ref{tab:multinest_screen_pixel} (see also Figs. \ref{fig:b_ne_rm_comparison_app_1} to \ref{fig:SLs_comparison_app_6} of the Appendix).

\subsection{Lens mass-density distribution}

From modelling the Stokes $\vec{I}$ emission, we recover the lens mass density distribution parameters. Fig. \ref{fig:mn_lens} shows the inferred posterior distribution for data sets D$_{\rm 6}$ and D$_{\rm 8}$, as an example. Table \ref{tab:multinest_mass} lists the mean parameter values with the corresponding 68 per cent confidence-level uncertainties for the same data sets. An equal level of constraint is obtained for all data sets with the same lensing configuration, as expected from their matching data quality (angular resolution and signal-to-noise ratio). 

\begin{table}
	\centering
	\caption{The recovered lens mass density parameters from modelling the Stokes $\vec{I}$ surface brightness distribution for data sets D$_{\rm 6}$ (source = $s_{\rm1}$, $P_f$ = 3 per cent, $i = 45$ deg) and D$_{\rm 8}$ (source = $s_{\rm2}$, $P_f$ = 3 per cent, $i = 45$ deg). We quote the mean values and the 68 per cent confidence limits for each parameter.}
	\label{tab:multinest_mass}
    \begin{tabular}{lll}
        \hline
        &D$_{\rm 6}$ & D$_{\rm 8}$\\
        \hline
        $\kappa_{\rm 0}$        & 0.60       $\pm$ 0.01       &  0.609 $\pm$ 0.006\\
	$\theta$ [deg]             & $-$34.4 $\pm$ 0.8         &  $-$32.1 $\pm$ 0.3\\  
        $q$                              & 0.839    $\pm$ 0.009       & 0.800 $\pm$ 0.003\\ 
	$\gamma$                   & 1.99      $\pm$ 0.02       & 2.019  $\pm$ 0.001\\
	
	$\kappa_{\rm 0,d}$      & 0.78      $\pm$ 0.02       & 0.779 $\pm$ 0.007\\ 
	$q_{\rm d}$                  & 0.13      $\pm$ 0.02       & 0.198 $\pm$ 0.002\\ 
	$R_{\rm d}$ [arcsec]    & 0.204    $\pm$ 0.004       & 0.185 $\pm$ 0.001\\   
	
	$\theta_{\rm d}$ [deg]  & $-$26.8 $\pm$ 0.3         & $-$31.6 $\pm$ 0.2\\ 
	$\Gamma_{\mathrm{sh}}$       & 0.035 $\pm$ 0.001       & 0.0393 $\pm$ 0.0005\\ 
	$\theta_{\mathrm{sh}}$ [deg]   & 13.6   $\pm$ 1.6  & 27.5 $\pm$ 0.5\\ 
        \hline
    \end{tabular}
\end{table}

\subsection{Rotation Measure}

\begin{figure*}
\centering
\includegraphics[scale=0.6]{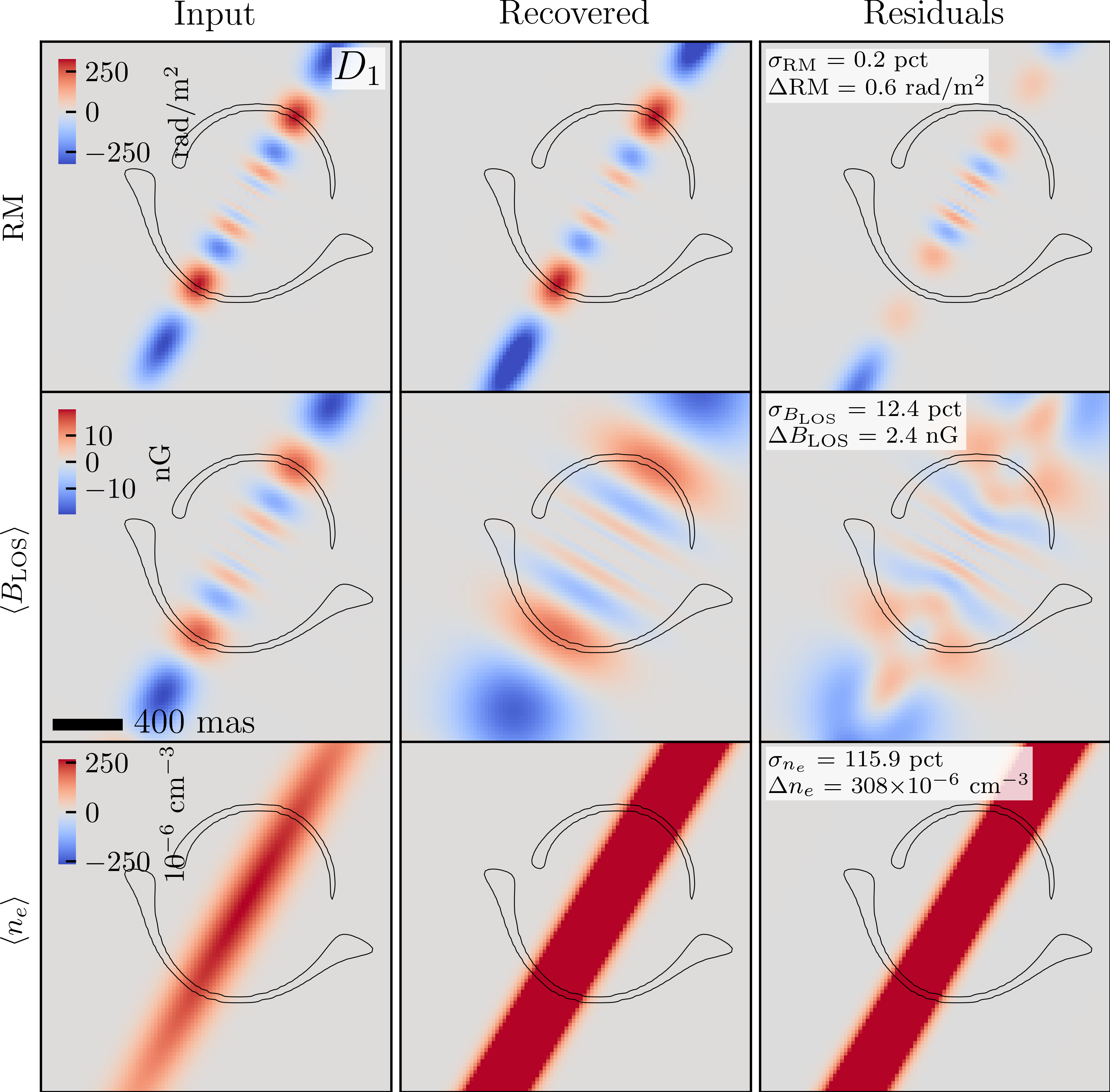}~~
\includegraphics[scale=0.6]{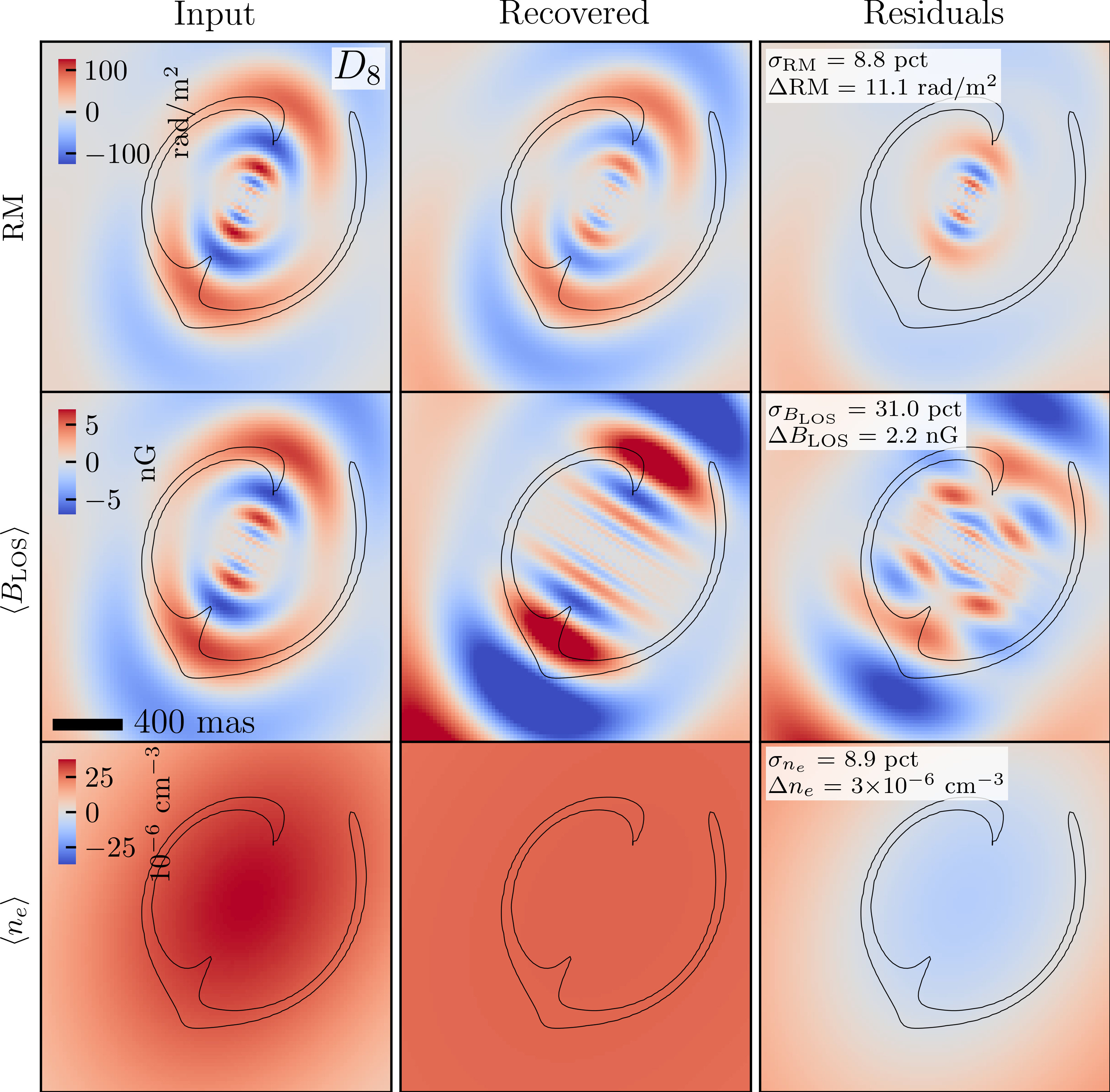}
\caption{Modelling results for D$_{\rm 1}$ (source = $s_{\rm1}$, $P_f$ = 12 per cent, $i$ = 90 deg; left) and D$_{\rm 8}$ (source = $s_{\rm2}$, $P_f$ = 3 per cent, $i$ = 45 deg; right) in the lowest frequency channel, as an example (note that we simultaneously model all channels). Panels contain the ground truth (left column), recovered (middle column), and residuals (right column) for the Rotation Measure RM (upper row), the projected line-of-sight magnetic field $\vec{B}_\mathrm{LOS}$ (middle row) and electron number density $\vec{n_{\rm e}}$ (lower row). In the residual column, we quote the fractional RMS difference between the input and recovered quantities, computed within the region where the lensed images are at least 20 per cent of the peak surface brightness (black contours). This illustrates the inherent degeneracies between $\vec{B}_\mathrm{LOS}$ and $\vec{n_{\rm e}}$. The RM is recovered to within a few per cent, but the uncertainties on $\vec{B}_\mathrm{LOS}$ and $\vec{n_{\rm e}}$ individually are much higher.}
\label{fig:b_ne_rm_comparison_d1d8}
\end{figure*}

\begin{figure*}
\centering
\includegraphics[scale=0.6]{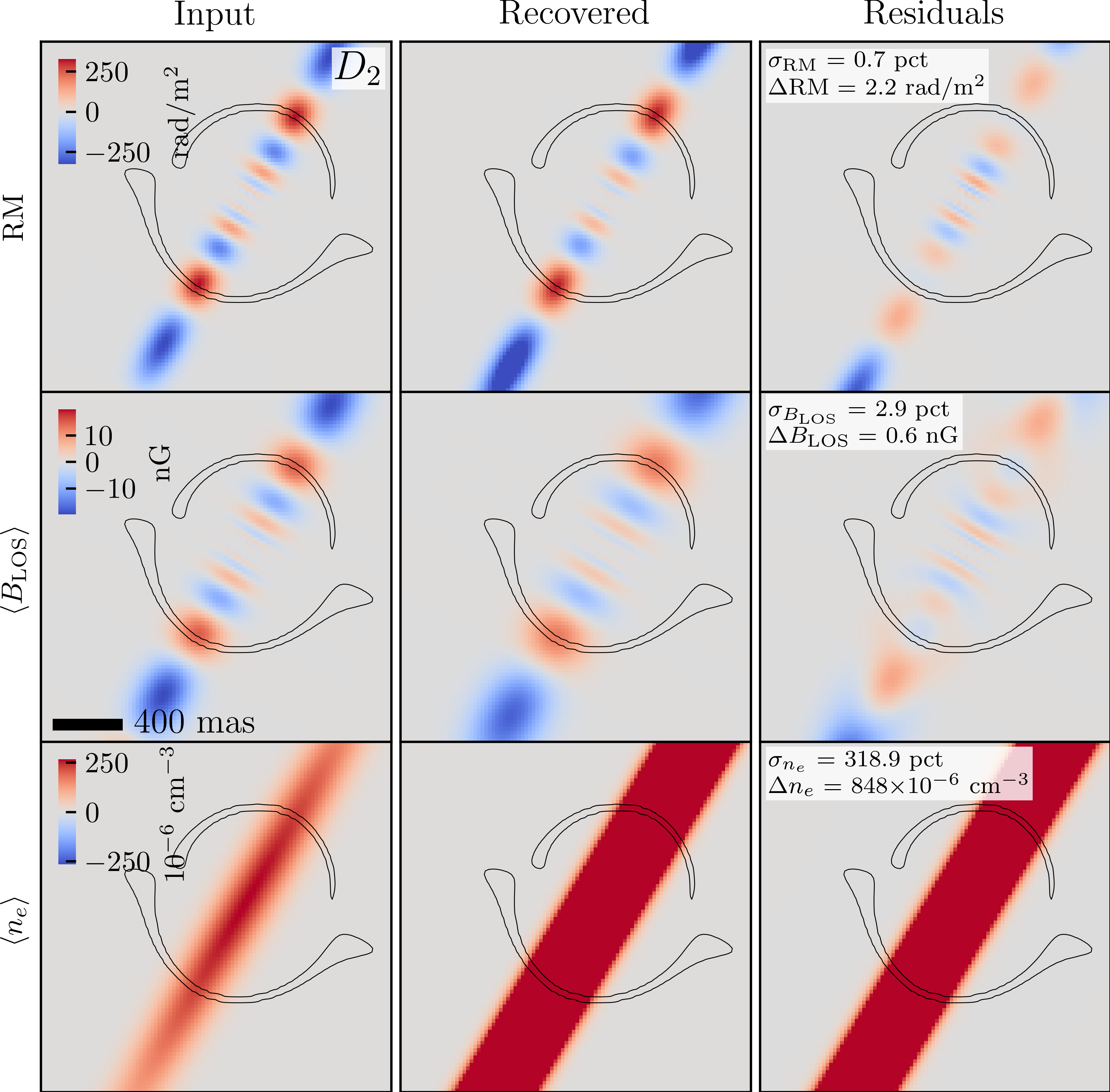}~~
\includegraphics[scale=0.6]{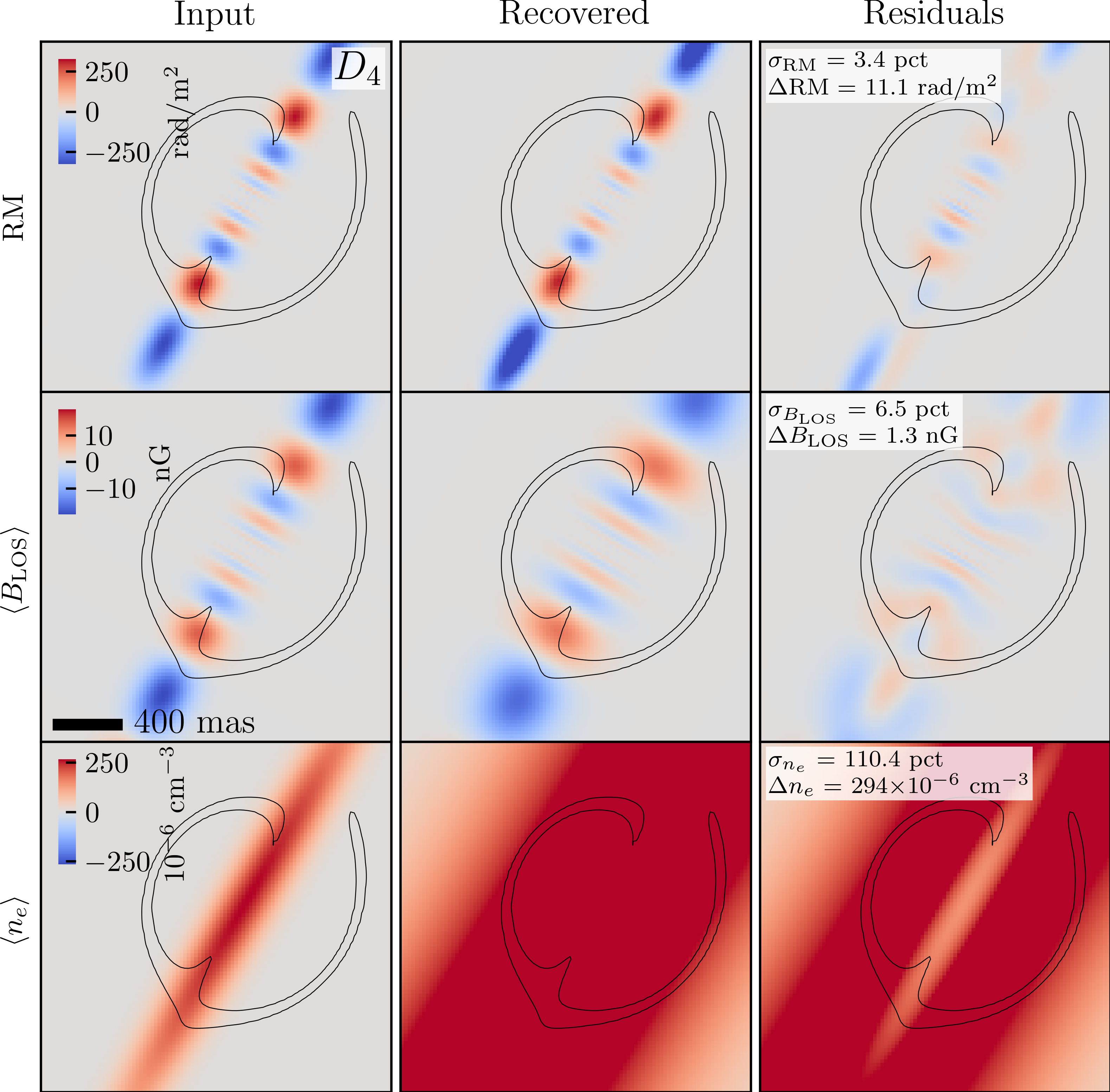}
\caption{Modelling results for D$_{\rm 2}$ (source = $s_{\rm1}$, $P_f$ = 3 per cent, $i$ = 90 deg; left) and D$_{\rm 4}$ (source = $s_{\rm2}$, $P_f$ = 3 per cent, $i$ = 90 deg; right) in the lowest frequency channel, as an example (note that we simultaneously model all channels). Panels contain the ground truth (left column), recovered (middle column), and residuals (right column) for the Rotational Measure RM (upper row), the projected line-of-sight magnetic field $\vec{B}_\mathrm{LOS}$ (middle row) and electron number density $\vec{n_{\rm e}}$ (lower row). In the residual column, we quote the fractional RMS difference between the input and recovered quantities, computed within the region where the lensed images are at least 20 per cent of the peak surface brightness (black contours).}
\label{fig:b_ne_rm_comparison_d2d4}
\end{figure*}

As indicated by the Bayes factor $\Delta\log{\cal E} = \log {\cal E}_{\rm screen} - \log {\cal E}_{\rm noscreen}$, the presence of a Faraday rotating screen in the lens galaxy is detected with a high statistical significance for all data sets (above 180$\sigma$) and even more so for those generated from the source $s_{\rm 2}$ (above 260$\sigma$; see Table \ref{tab:multinest_screen_pixel}). The latter result is expected because a larger source probes a larger fraction of the lens plane. At a fixed polarisation fraction, we find that different lensing configurations are better at probing different RM distributions: the screen is more strongly detected when it has an inclination angle of 90 and 45 deg for $s_{\rm 1}$ and  $s_{\rm 2}$, respectively. This is due to the more localised effect of the 90-deg inclination model. The source $s_{\rm 2}$, unlike $s_{\rm 1}$, leads to a counter-image that only marginally overlaps with a positive peak in the RM distribution. On the other hand, the 45-deg inclination model is more extended and hence is more strongly detected with a larger source.

Fig. \ref{fig:b_ne_rm_comparison_d1d8} displays our best (D$_{\rm 1}$; $\sigma_{\rm RM}$ = 0.2 per cent, $\Delta{\rm RM}$ = 0.6 rad~m$^{-2}$) and worst (D$_{\rm 8}$; $\sigma_{\rm RM}$ = 8.8 per cent, $\Delta{\rm RM}$ = 11.1 rad~m$^{-2}$) results, respectively. It can be seen that even in the best-case scenario, where the RM is reconstructed with high accuracy at the location of the lensed images, departures from the truth are possible outside this area, indicating, as expected, that we have no constraint there. These results highlight the advantage of using resolved lensed sources over unresolved ones. In general, a combination of the underlying true RM distribution and the lensing configuration sets the constraints on the recovered RM. For example, data sets D$_2$ and D$_4$ are generated with the same level of source polarisation fraction and share the same underlying model for $\vec{B}$ and $\vec{n_{\rm e}}$. However, the former leads to better results ($\sigma_{\rm RM}$ = 0.7 per cent, $\Delta{\rm RM}$ = 2.2 rad~m$^{-2}$ versus $\sigma_{\rm RM}$ = 3.4 per cent, $\Delta{\rm RM}$ = 11.1 rad~m$^{-2}$) because of the different locations of the counter image with respect to the relevant features in the RM map (see Fig. \ref{fig:b_ne_rm_comparison_d2d4}). The same effect explains the relative performance between D$_3$ ($\sigma_{\rm RM}$ = 2.3 per cent, $\Delta{\rm RM}$ = 7.5 rad~m$^{-2}$) and D$_{\rm 1}$ ($\sigma_{\rm RM}$ = 0.2 per cent, $\Delta{\rm RM}$ = 0.6 rad~m$^{-2}$). On the other hand D$_{\rm 5}$ ($\sigma_{\rm RM}$ = 2.6 per cent, $\Delta{\rm RM}$ = 3.3 rad~m$^{-2}$) and D$_{\rm 7}$  ($\sigma_{\rm RM}$ = 2.3 per cent, $\Delta{\rm RM}$ = 3.0 rad~m$^{-2}$),  and D$_{\rm 6}$  ($\sigma_{\rm RM}$ = 8.8 per cent, $\Delta{\rm RM}$ = 11.1 rad~m$^{-2}$) and D$_{\rm 8}$  ($\sigma_{\rm RM}$ = 8.8 per cent, $\Delta{\rm RM}$ = 11.1 rad~m$^{-2}$) lead to a similar level of constraints because the less localised RM structure of the 45-degree inclination angle model can be probed equally well by both lensing configurations.

At a fixed inclination angle and lensing configuration, data sets with a larger polarisation fraction lead to better results due to their higher signal-to-noise ratio, as expected.

\begin{table*}
	\centering
	\caption{Faraday screen parameters recovered from modelling all Stokes parameters. Upper: the mean values and 68 CL uncertainty for the magnetic field model, with $A_{\rm 0}= B_{\rm 0}\times n_{\rm 0}$. The last column is the difference in Bayesian Evidence relative to a model without a Faraday screen. Lower: the mean values and 68 CL uncertainty for the electron density model.}
	\label{tab:multinest_screen_pixel}
	\begin{tabular}{l l l l l l l l l l}
 \hline
   \multicolumn{10}{c}{Recovered magnetic field parameters}\\
   \\
   Data set & $A_{\rm 0}$ & $h$ & $r_{\rm 0}$ &  $\phi_{\rm 0}$ & $p$ & $\chi_{\rm 0}$ & $i$ & $\theta$ & $\Delta\log {\cal E}$\\
   & [$\mu$G~cm$^{-3}$] & [kpc] & [kpc] & [deg]  & [deg] & [deg] & [deg] & [deg]\\
    \hline
    D$_{\rm 1}$ & 0.25 $\pm$ 0.01 & 1.8 $\pm$ 0.2 & 13 $\pm$ 3 & 150 $\pm$ 7  & 11.8 $\pm$ 0.2
    & 38 $\pm$ 7 & 89.9 $\pm$ 0.5 & $-$30.0 $\pm$ 0.1& 17162\\
    D$_{\rm 2}$ & 0.24 $\pm$ 0.03 & 1.4 $\pm$ 0.3 & 14 $\pm$ 5 & 133 $\pm$ 18 & 11.6 $\pm$ 0.5 & 38 $\pm$ 6 & 89.9 $\pm$ 0.8 & $-$30.1 $\pm$ 0.3 & 18670\\
    D$_{\rm 3}$ & 0.228 $\pm$ 0.002 & 2.0 $\pm$ 0.3 & 23.0 $\pm$ 0.8 & 98 $\pm$ 3 & 10.38 $\pm$ 0.07 &
    37 $\pm$ 7 & 89.25 $\pm$ 0.03 & $-$30.32 $\pm$ 0.04 & 34463 \\
    D$_{\rm 4}$ & 0.24 $\pm$ 0.03 & 1.5 $\pm$ 0.3 & 20 $\pm$ 11 & 130 $\pm$ 9 & 11.2 $\pm$ 0.2 & 37 $\pm$ 7 & 88 $\pm$ 1 & $-$30.6 $\pm$ 0.3 & 36192 \\
    D$_{\rm 5}$ & 0.24 $\pm$ 0.02 & 1.7 $\pm$  0.2 & 12 $\pm$ 2 & 11 $\pm$ 7 & 8.6 $\pm$ 0.1 & 43 $\pm$ 3 & 43.7 $\pm$ 0.3 & $-$31.2 $\pm$ 0.3 & 18263\\
    D$_{\rm 6}$ & 0.18 $\pm$ 0.01 & 3.6 $\pm$ 0.9 & 18 $\pm$ 3 & 7 $\pm$ 13 & 8.1 $\pm$ 0.2 & 42 $\pm$ 4 & 39.6$\pm$ 0.9 & $-$45 $\pm$ 1 & 18839\\
    D$_{\rm 7}$ & 0.29 $\pm$ 0.01 & 1.6 $\pm$ 0.1 & 7.5 $\pm$ 0.4 & 69 $\pm$ 2 & 9.74 $\pm$ 0.04 & 44 $\pm$ 3 & 45.6 $\pm$ 0.1 & $-$30.9 $\pm$ 0.1 & 34703\\
    D$_{\rm 8}$ & 0.20 $\pm$ 0.01 & 3.4 $\pm$ 1.0 & 19 $\pm$ 3 & 30 $\pm$ 8 & 8.9 $\pm$ 0.1 & 42 $\pm$ 4 & 46.0 $\pm$ 0.4 & $-$32.5 $\pm$ 0.4 & 36149\\
    \hline 
  \multicolumn{10}{c}{Recovered electron density parameters}\\
  \\
  Data set &  $h_{\rm ne}$ & $r_{\rm ne}$\\
    &  [kpc] & [kpc] \\
    \hline
    D$_{\rm 1}$ & 0.76 $\pm$ 0.02 & 84 $\pm$ 35\\
    D$_{\rm 2}$ & 0.83 $\pm$ 0.09 & 79 $\pm$ 38\\
    D$_{\rm 3}$ & 0.71 $\pm$ 0.01 & 144 $\pm$ 10\\
    D$_{\rm 4}$ & 0.74 $\pm$ 0.05 & 99 $\pm$ 31\\
    D$_{\rm 5}$ & 0.91 $\pm$ 0.04 & 31 $\pm$ 10\\
    D$_{\rm 6}$ & 1.01 $\pm$ 0.05 & 34 $\pm$ 11\\
    D$_{\rm 7}$ & 0.83 $\pm$ 0.03 & 41 $\pm$ 8\\
    D$_{\rm 8}$ &0.71  $\pm$ 0.04 & 39 $\pm$ 9\\
     \hline 
    \end{tabular}
\end{table*}

\begin{figure*}
\centering
\includegraphics[angle= 90,width=0.9\textwidth]{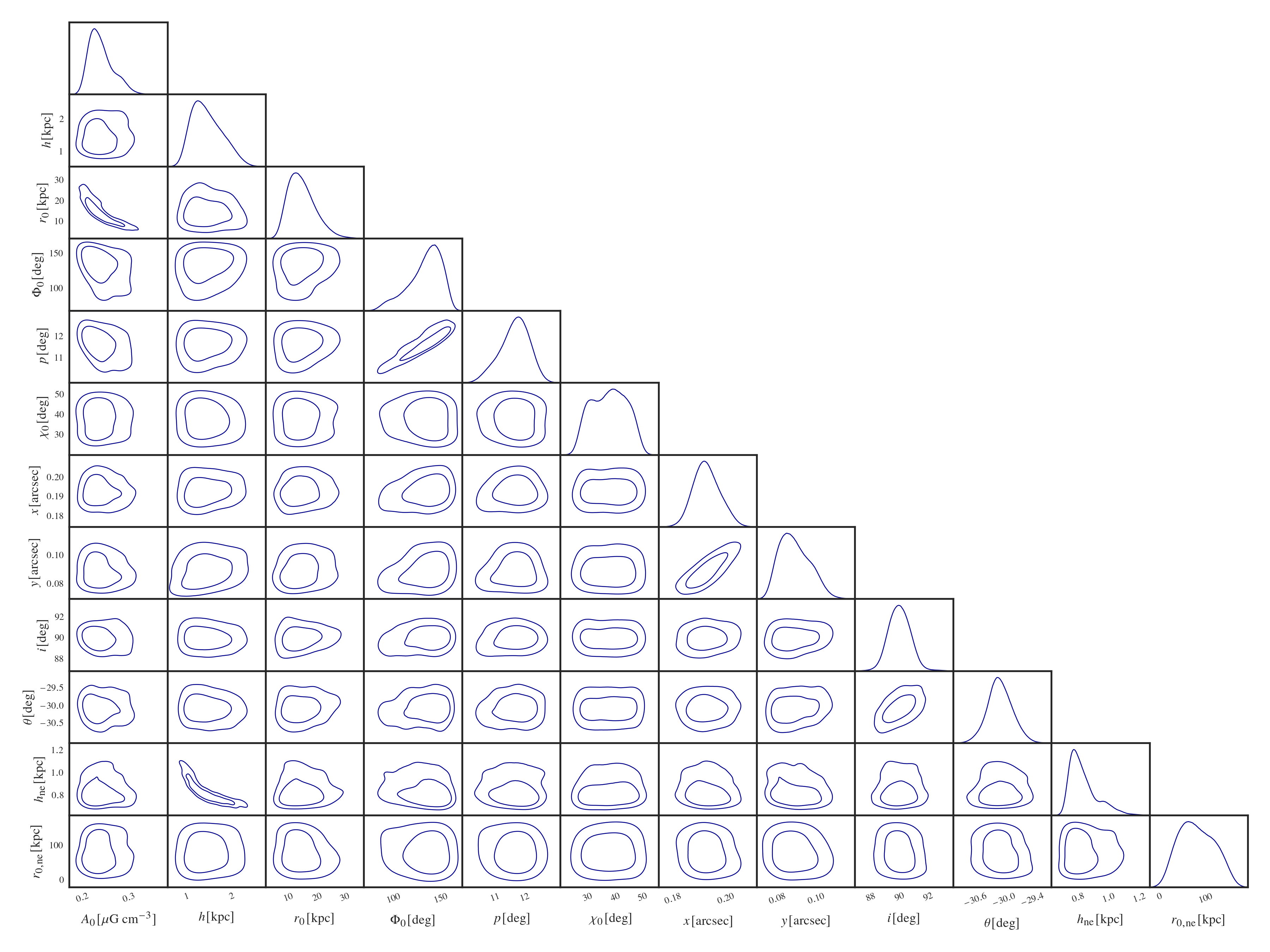}
\caption{The posterior distribution for all parameters of the lens magnetic field and electron density models for data set D$_{\rm 2}$ (source = $s_{\rm 1}$, $P_f$ = 3 per cent, $i = 90$ deg).}
\label{fig:mn_d2}
\end{figure*}

\begin{figure*}
\centering
\includegraphics[angle=90,width=0.9\textwidth]{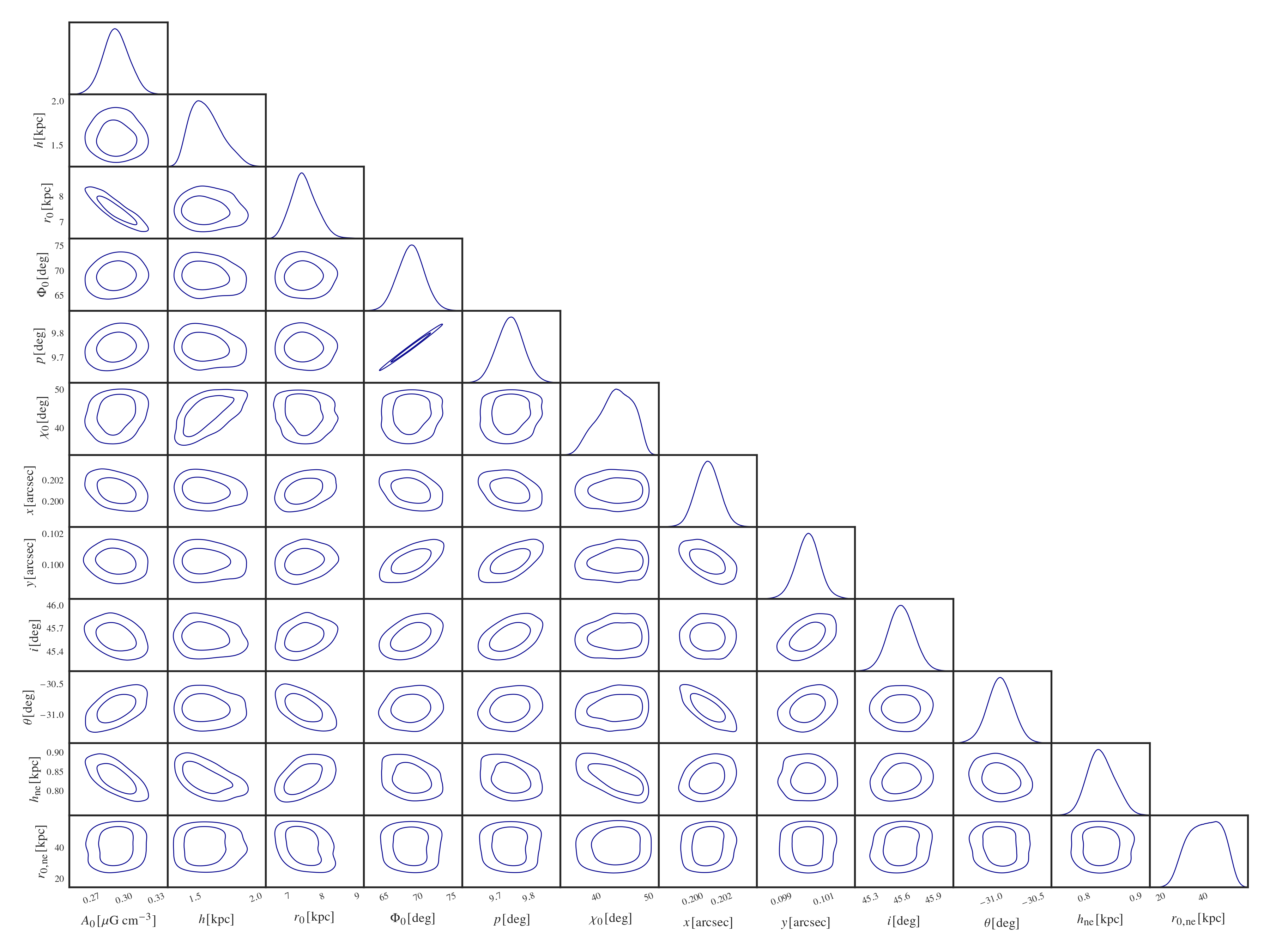}
\caption{The posterior distribution for all parameters of the lens magnetic field and electron density model for data set D$_{\rm 7}$ (source = $s_{\rm 2}$, $P_f$ = 12 per cent, $i = 45$ deg).}
\label{fig:mn_d7}
\end{figure*}

\subsection{Magnetic field and electron density profiles}

We recover the electron density profile with an accuracy between $\sigma_{\rm n_e}$ = 5.4 per cent and $\Delta{\rm n_e} = 2.0\times10^{-6}$~cm$^{-3}$ in the best case (for data set D$_{\rm 5}$), and $\sigma_{\rm n_e}$ = 319 per cent and $\Delta{\rm n_e} = 8.48\times10^{-4}$~cm$^{-3}$ in the worst case (for data set D$_{\rm 2}$). Our choice of a Gaussian profile with a scale height that is almost twice as large as the Einstein radius essentially leads to a constant value of the electron density at the location of the lensed images. As a consequence, we find this quantity to be largely insensitive to the lensing configuration and, hence, will likely not be well constrained in general by galaxy-scale gravitational lensing with typical Einstein radii of 0.15 to 1.5 arcsec (or 0.5 to 5 kpc in linear size, for a lens redshift of 0.5). On the other hand, we see a clear trend with the inclination angle; data sets with $i=45$~deg perform consistently better when compared to those with $i=90$~deg. This is due to the magnetic field profile sharply dropping to zero along the vertical direction in the first model and from the degeneracy between the two scale heights $h_{\rm ne}$ and $h$. For an inclination angle of $i=45$ deg and at a fixed lensing configuration, data sets with a higher signal-to-noise ratio (i.e. a larger polarisation fraction) provide better constraints.

For the line-of-sight magnetic field profile, $\vec{B}_{\rm LOS}$, we recover the input model within the region probed by the lensed images with an accuracy that ranges from $\sigma_{\rm B_{LOS}}$ = 2.9 per cent and $\Delta{\rm B_{LOS}}$ = 0.3 nG to $\sigma_{\rm B_{LOS}}$ = 151 per cent and $\Delta{\rm B_{LOS}}$ = 3.0 nG. Similar to the RM profile, the lensing configuration plays a substantial role in the 90-deg inclination angle model, but less so for the 45-deg case. This aligns with the fact that structures in the RM profile come from the magnetic field model and are just rescaled by the almost constant electron density model. For both inclination angles, data sets with a larger polarisation fraction lead to a higher accuracy for a given lensing configuration. As above, we find that reliable constraints can only be obtained within the region probed by the lensed images.

We note that a good accuracy in the RM model does not necessarily translate to good quality constraints for the two separate quantities of $\vec{n_{\rm e}}$ and  $\vec{B}_{\rm LOS}$,  which is due to the inherent degeneracy arising from equation (\ref{eq:faraday_depth}). In this respect, the magnetic field and electron density models are \emph{latent parameters} that are not directly measurable. Hence, one should be careful when interpreting the inferred RM in the context of real data. This issue is likely to be model-dependent, and we will investigate it more broadly in a follow-up study.

\begin{figure*}
\centering
\includegraphics[scale=0.6]{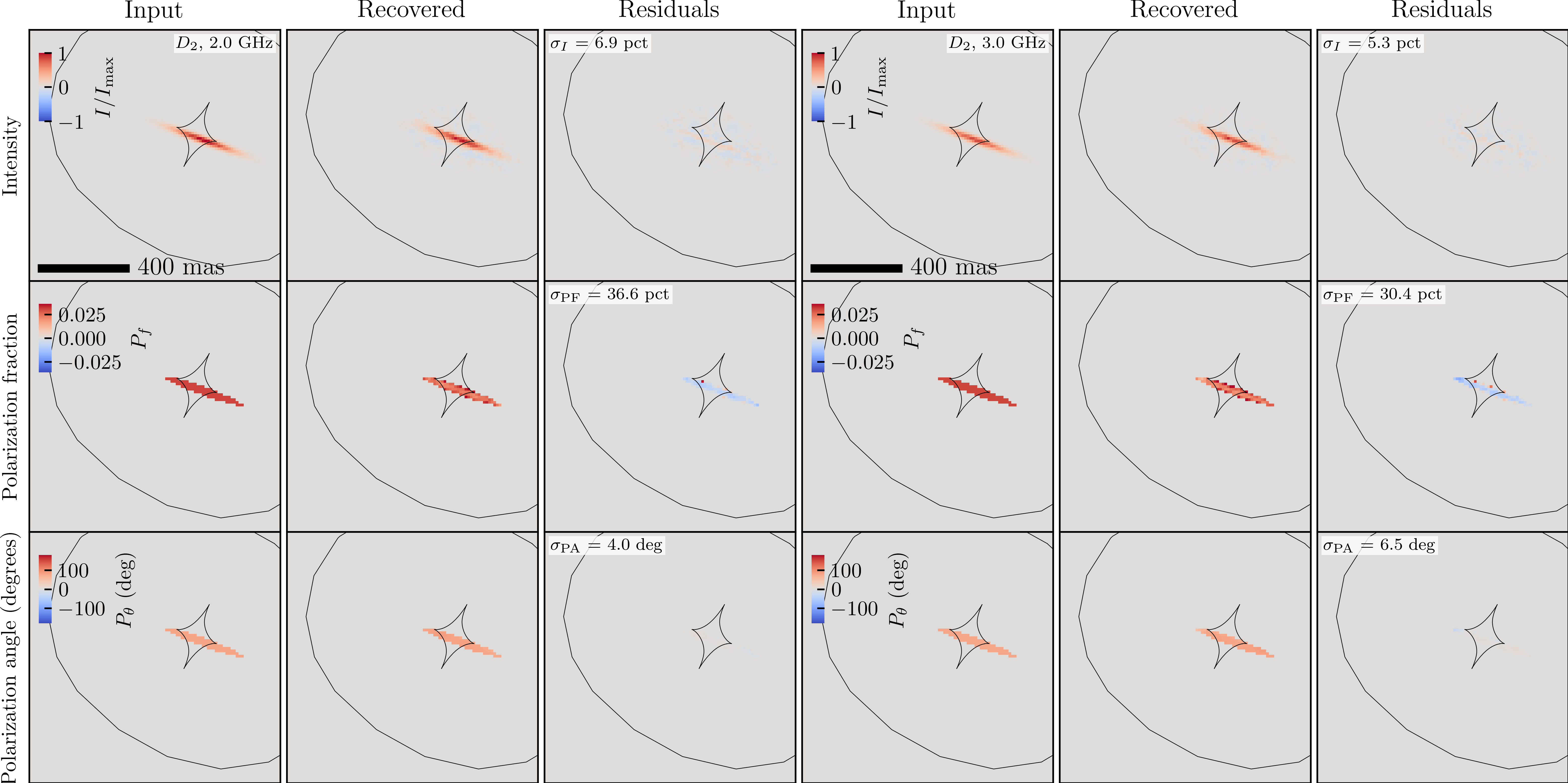}\\~\\
\includegraphics[scale=0.6]{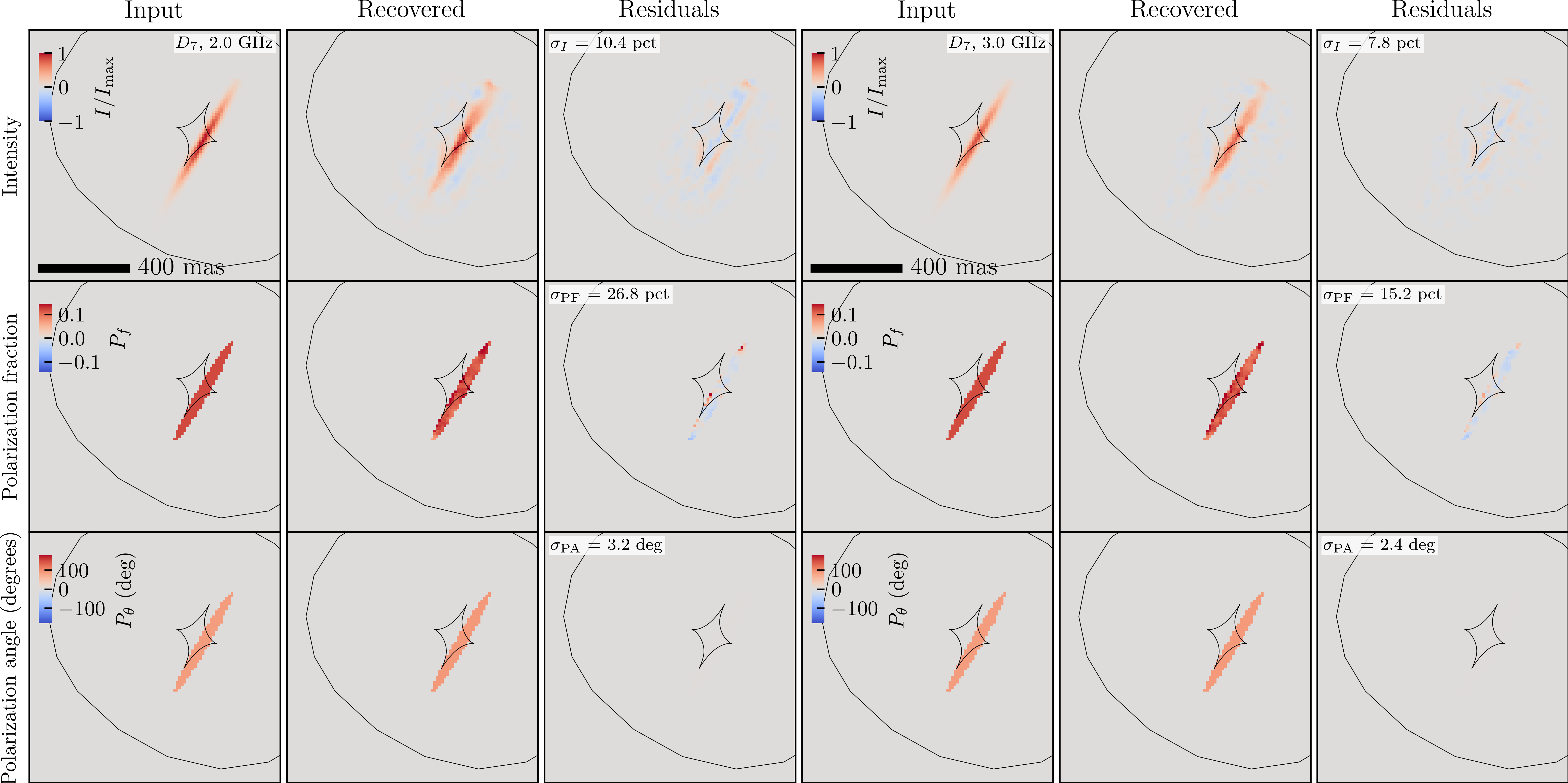}
\caption{Comparison of the recovered source-plane results with respect to the input model, for D$_{\rm 2}$ (source = $s_{\rm1}$, $P_f$ = 3~per cent, $i$ = 90 deg; top) and D$_{\rm 7}$ (source = $s_{\rm2}$, $P_f$ = 12~deg, $i$ = 45 deg; bottom) at 2 and 3 GHz, as an example (note that we simultaneously model all channels). The rows contain the Stokes $\vec{I}$ (top), the polarisation fraction (middle), and the polarisation angle (in deg; bottom) for both data sets.  For clarity, we mask the colour maps to include only the region of the highest SNR, where the input surface brightness Stokes $\vec{I}$ is at least 20 per cent of the peak. The residual RMS values are computed within the same mask.}
\label{fig:src_comparison}
\end{figure*}

\begin{figure*}
\centering
\includegraphics[scale=0.6]{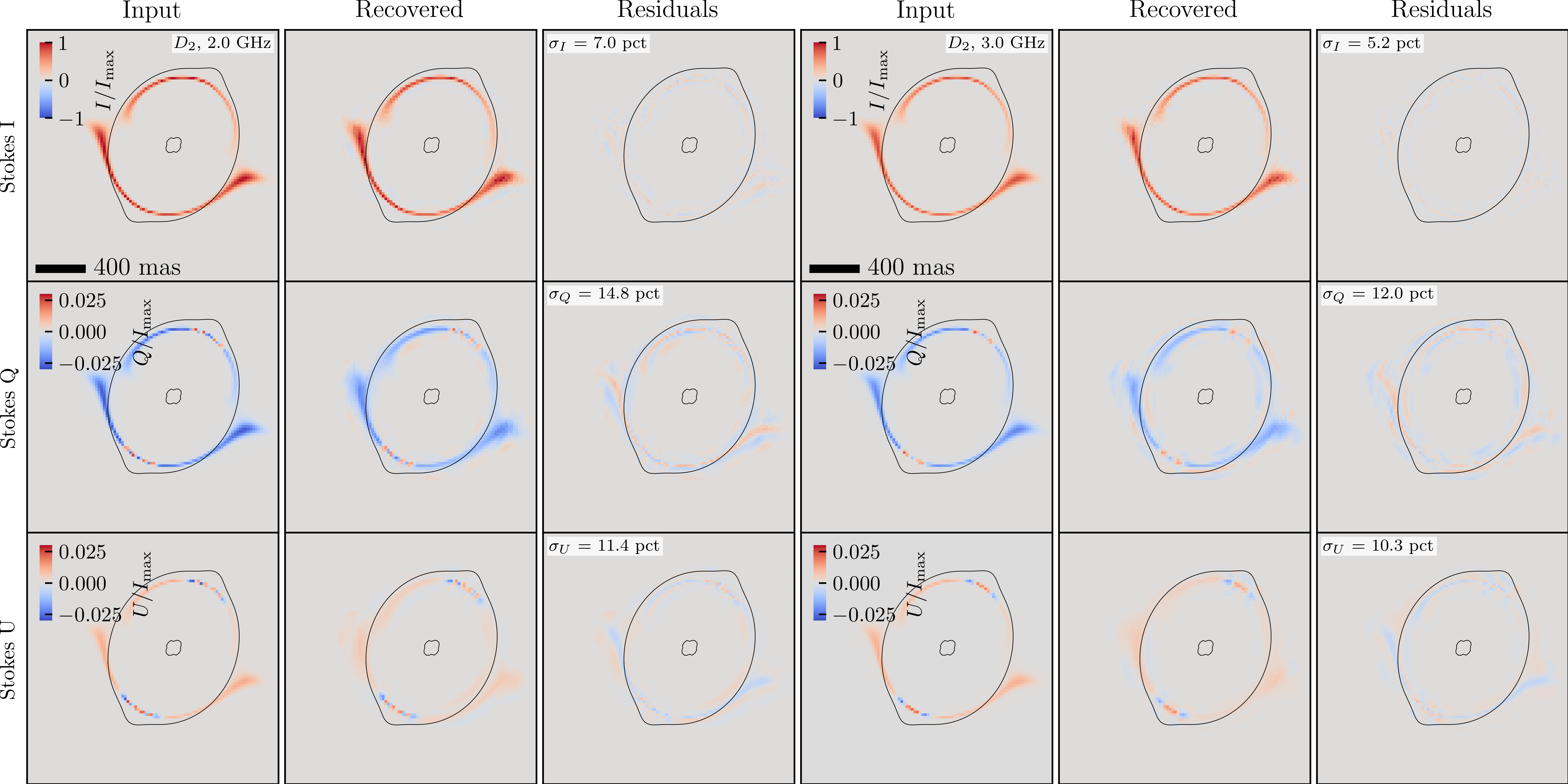}\\~\\
\includegraphics[scale=0.6]{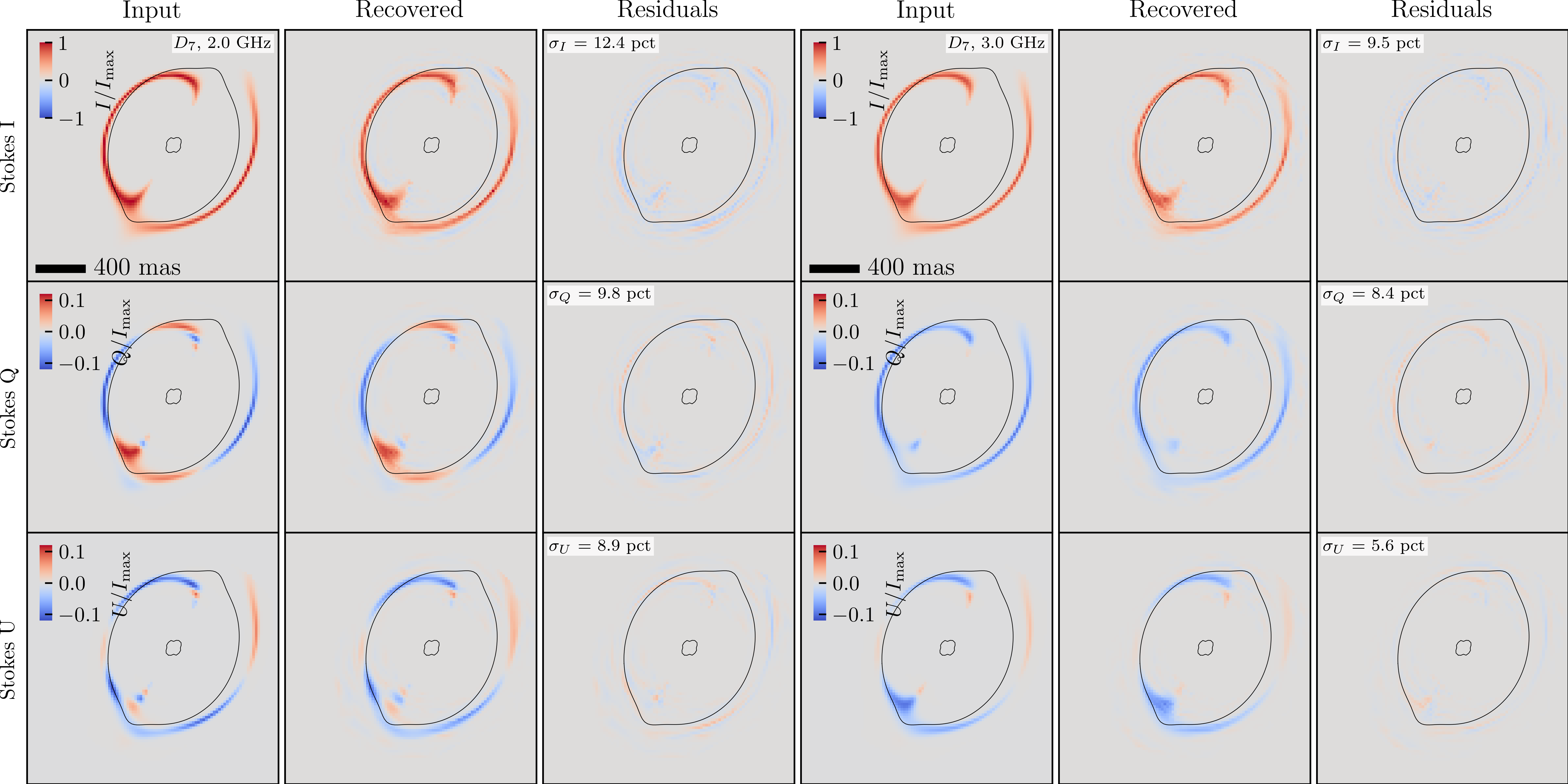}
\caption{Comparison of the recovered image-plane results with respect to the input model, for D$_{\rm 2}$ (source = $s_{\rm1}$, $P_f$ = 3~per cent, $i$ = 90 deg; top) and D$_{\rm 7}$ (source = $s_{\rm2}$, $P_f$ = 12~per cent, $i$ = 45 deg; bottom) at 2 and 3 GHz, as an example (note that we simultaneously model all channels). The rows contain Stokes $\vec{I}$, $\vec{Q}$, and $\vec{U}$ from top to bottom. We do not show Stokes $\vec{V}$, but find results consistent with zero surface brightness, in agreement with our input model. For clarity, we mask the colour maps to include only the region of the highest SNR, where the input surface brightness Stokes $\vec{I}$ is at least 20 per cent of the peak. The residual RMS values are computed within the same mask.}
\label{fig:SLs_comparison}
\end{figure*}

\subsection{Magnetic field and electron density model degeneracies}

In Table~\ref{tab:multinest_screen_pixel}, we give the mean values and the 68 per cent confidence-level uncertainties derived with \software{MultiNest} for all magnetic field and electron density model parameters. The corresponding posterior distributions for data sets D$_{\rm 2}$ and D$_{\rm 7}$, our best-recovered models for a disk inclination of respectively 90 and 45 deg, are presented in Figs.~\ref{fig:mn_d2} and \ref{fig:mn_d7}, with all remaining data sets presented in Figs. \ref{fig:mn_90deg} and \ref{fig:mn_45deg} of the Appendix.

From the posterior distributions, we see that all the expected degeneracies are recovered. In the case of the RM, which is an integrated quantity, we find that the parameter $A_{\rm 0}$ is underestimated by, on average, 30~per cent for all data sets and by a factor of 2 in the worst case (for data set D$_{\rm 6}$). This is due to $r_{\rm_0}$ and $h$ of the magnetic field model being over-estimated by, on average, a factor of 1.6 and 2, respectively, and given the degeneracy we see between $A_{\rm 0}$ and $r_{\rm_0}$, as opposed to an issue with the recovered RM (e.g. see Fig.~\ref{fig:b_ne_rm_comparison_d1d8}). 

The relative spacing between the positive and negative RM features is affected by both the pitch $p$ and phase $\phi_{\rm 0}$ angles. While the former is relatively well constrained within about 2 deg from the truth, the latter can be off by up to 70~deg. For the 90-deg inclination angle, we also recover the degeneracy between the magnetic field and electron density scale heights. The latter is underestimated by a factor of 1.4. As discussed already above, the input scale radius of the electron density distribution is about twice as large as the Einstein radius of the lens model, which in combination with our adopted model for $\vec{n_{\rm e}}$, makes our simulated data sets insensitive to values of $r_{\rm ne} > R_{\rm ein}$; any value above this limit results in the same Faraday rotating screen within the region where it is measured. As a result, we significantly overestimate this parameter by up to a factor of 15. Both the inclination and position angles are well recovered, with at most a difference of 5 to 15 deg from the input model, respectively. As expected, they are both more tightly constrained for inclination angles of 90~deg due to the stronger and much more localised effect of the screen when compared to the 45-deg case. The limit on the tilt angle $\chi_{\rm 0}$ is at most 8 deg off from the input value.

We can conclude that the accuracy with which each parameter is inferred depends on the many factors at play, namely the lensing configuration and how it probes different inclination angle models, the signal-to-noise ratio of the data and the intrinsic degeneracies, not only between the electron density and the magnetic field models, but also between the parameters within each model. For example, we find that data set D$_{\rm 7}$ is the one that performs best in terms of recovering the largest number of parameters with best accuracy (between 2 and 13 per cent). This is likely related to the low inclination angle and the high signal-to-noise ratio of that data set. On the other hand, D$_{\rm 6}$ provides the worst constraints for most parameters, for example, by underestimating $A_{\rm 0}$ by almost a factor of 2 and overestimating $h$ by nearly a factor of 4. This particular data set has the smallest source size and polarisation fraction and, therefore, provides the worst constraints in terms of the structure of the lensed images and their signal-to-noise ratio. Thus, although the RM can be well reproduced with this method, the individual model parameters of the electron density and magnetic field models may not be without additional information that can break some of the degeneracies at play.

\subsection{Background source}

We now consider the ability of our method to recover the {\it intrinsic} properties of the background source, namely the recovered surface brightness distribution, polarisation fraction and polarisation angle. We re-iterate here that although we have used a smooth parametric model to describe the background source surface brightness distribution in the Stokes $\vec{I}$, $\vec{Q}$ and $\vec{U}$, our recovered model has no prior information on this, but is instead based on a regularised pixellated model. The recovered source can be used to infer the robustness of the method since the presence of an intervening magneto-ionic medium will result in a differential Faraday rotation of the data, which will produce a change in the surface brightness in the Stokes $\vec{Q}$ and $\vec{U}$ that is inconsistent with lensing. This will result in a mismatch between the observed and modelled surface brightness in the lensed emission, which will be compensated for with an inaccurate model for the background source. Also, measuring the polarisation properties of the gravitationally lensed background source will be in itself interesting for resolving the magnetic field distribution within high redshift sources (e.g. \citealt{Geach2023}), which is applicable whether there is an effective intervening magneto-ionic medium (low frequencies; LOFAR, SKA and ngVLA) or not (high frequencies; ALMA).

Fig. \ref{fig:src_comparison} presents the input and recovered source surface brightness distribution for Stokes $\vec{I}$, and the polarisation fraction and polarisation angle for the data sets D$_{\rm 2}$ and D$_{\rm 7}$ at 2 and 3 GHz, as an example (results for the other data sets can be found in Figs. \ref{fig:SLs_comparison_app_1} to \ref{fig:SLs_comparison_app_6} of the Appendix). The corresponding lensed surface brightness distributions for Stokes $\vec{I}$, $\vec{Q}$ and $\vec{U}$ can be seen in Fig. \ref{fig:SLs_comparison}. We do not show the results for Stokes $\vec{V}$, but we find it consistent with zero across all frequency channels within the noise. We see from inspecting the residuals presented in Fig. \ref{fig:src_comparison} that the method recovers the input source surface brightness very well for both data sets due to the high signal-to-noise ratio of the total intensity emission. We find that the absolute polarisation fraction and angle are recovered to within 40 and 14 per cent and to within 4 and 3 deg across the recovered source for $s_{\rm 1}$ and $s_{\rm 2}$, respectively. The most significant deviations in the polarisation fraction are seen for data set  D$_{\rm 2}$, which has a factor of 4 lower signal-to-noise ratio in polarised flux. This again demonstrates how the robustness of the results will be affected by the quality of the data.

We have also calculated the source residual RMS within the region of the highest signal-to-noise ratio, where the input Stokes $\vec{I}$ surface brightness is at least 20 per cent of the peak. As the total intensity is not affected by the Faraday screen, we find that the quality of the reconstruction solely depends on the lensing configuration, with a relative RMS of about 7 and 10 per cent for $s_{\rm 1}$ and $s_{\rm 2}$, respectively. This is likely related to the fact that a larger fraction of $s_{\rm 1}$ is within the quadruply-imaged region, providing more constraints for the source reconstruction. The same effect is observed in Fig. \ref{fig:SLs_comparison} for Stokes $\vec{Q}$ and $\vec{U}$ at fixed polarisation fraction and for the inclination angle of the underlying RM model. A higher level of polarisation fraction leads to a higher signal-to-noise ratio and, hence, lower RMS residuals for Stokes $\vec{Q}$ and, in particular, $\vec{U}$. We note that for a fixed lensing configuration and polarisation fraction, the Stokes $\vec{Q}$ surface brightness is better recovered when the underlying RM model has a lower inclination angle.

\section{Discussion}
\label{sec:limitation}

In this section, we discuss the prospects of our method for measuring the magneto-ionic medium of intermediate redshift galaxies and the future developments that we plan to carry out.

\subsection{Probing the magnetic field structure of intermediate redshift galaxies}

We have developed a method for the gravitational lens modelling of complex polarised sources at radio to mm-wavelengths, with the aim of constraining models for the magneto-ionic medium of the foreground lens. This important component of galaxies is extremely challenging to constrain, except for the very nearby Universe, where the resolved polarised emission from massive star-forming galaxies can be mapped \citep{Fletcher2011,Heald2016,Mulcahy2017}. However, it has been recently shown that gravitational lens systems with a simple point-like lensed source can provide a new channel for such studies at intermediate redshifts \citep{Mao2017}. Here, we have developed the first gravitational lens modelling code that can simultaneously model the lensing mass distribution, the effect of an intervening magneto-ionic medium and recover the polarised properties of the background source, all in the native visibility plane of the data.

The key observable when it comes to polarised emission is the Stokes $\vec{Q}$ and $\vec{U}$ surface brightness, from which the electric-vector polarisation angle can be determined, which if known as a function of frequency allows the RM to be measured; it is this measurement that links the observed emission to the astrophysics via the dependence of the magnetic field strength and the electron density \citep{Reissl2023}. Indeed, from our simulations, we have shown that the observed surface brightness of the lensed images probes different parts of the foreground lens galaxy within a few to a few tens of kpc in projection, allowing a relative change in the magnetic field structure to be detected. Since gravitational lensing conserves the surface brightness of the lensed source, any change in the polarisation angle that is seen in the lensed emission must be due to the presence of a magneto-ionic medium along the line-of-sight (in the absence of variability or an axion-like dark matter particle; see below). This is found at high confidence from our simulations when we compared the results from modelling Faraday rotated data with and without the inclusion of a Faraday screen. 

Therefore, this methodology, when applied to a large sample of polarised sources that are gravitationally lensed, can potentially detect evidence of such an intervening medium in the first instance and provide a method for selecting and modelling those objects of interest for detailed follow-up at higher signal-to-noise ratio and angular resolution. This will be particularly relevant for the large-area surveys to be carried out with, for example, LOFAR2.0 and SKA-MID in the near-term \citep{McKean2015}. These instruments are expected to find a large population of lensed radio sources that are also polarised. However, in addition to detecting evidence for a magneto-ionic medium, we also find that the RM at the location of the lensed emission is well-recovered, with an average error of around 0.6 to 11 rad~m$^{-2}$; this is an interesting result because, not only is it precise, it has been determined through modelling the magneto-ionic medium and the extended lensed emission simultaneously, as opposed to making independent measurements from the different lensed images to infer the former (e.g., \citealt{Mao2017}).

Our ability to recover the underlying structure of the magneto-ionic medium and separate the contributions of the magnetic field and the electron density profiles was more challenging than measuring the RM. This is a general problem for polarisation studies that also use non-lensed objects \citep{O'Sullivan2023} as there is an inherent degeneracy between these two components. However, we found that the product of the magnetic field strength and the electron density normalisation could be recovered to within a factor of two, which is extremely encouraging given that there is no other method that can probe the inner parts of such galaxies at intermediate redshifts. 

We do find that the RM profile can be measured at the location of the lensed emission, but due to the similar effect that various combinations of magnetic field structure can have, different models can likely well-fit the data. This is because the lensed emission still only probes a limited area of the galaxy structure, and since the lens magnification is mainly in a tangential direction (for a close to isothermal mass distribution), there is only limited information in the radial direction for a single-component polarised source. Therefore, the angular scale of any detectable variation in the magnetic field profile will be limited by the resolution of the data and the configuration of the lensed emission. That said, for the models tested here, we do find that the properties of the magnetic field are relatively well-recovered. This motivates further investigation into what types of RM profiles can be constrained for a wide range of magnetic field structures and lensing configurations. This will be the subject of future work. However, we note that combining with other data, for example, deep optical and infrared imaging, which can be used as a prior on the ellipticity and position angle of a magneto-ionic disk or its structure via the observed morphology of the spiral arms, we may be able to break some of the degeneracies.

Finally, we have also developed a method for recovering the underlying polarised properties of the lensed source, corrected for any propagation effects as the emission passes through the magneto-ionic medium of the foreground lens. Overall, the surface brightness distribution of the lensed source (its size, ellipticity and position angle) are well recovered since we take advantage of the Stokes $\vec{I}$ emission to simultaneously constrain the lens model parameters. This means that the polarised emission is also well-recovered because any uncertainties due to the lens model parameters are minimised. We find that for a good signal-to-noise ratio, the polarisation fraction and the electric-vector polarisation angle of the background source are recovered to about 15 per cent and 1 deg, respectively, from the ground truth. The similarity of the polarised source with our input model also demonstrates how well the foreground magento-ionic medium in the lens has been modelled with our method. We note that our method can also be applied to study high redshift galaxies at mm-wavelengths where propagation effects within the lens are thought to be negligible (e.g., \citealt{Geach2023}).

\subsection{Future developments and prospects}

Our current implementation of the method presented here is just the first step. There are still a number of improvements that can be made. For example, the method does not include the contribution of small-scale random magnetic fields, which are expected to dominate the interstellar medium of elliptical galaxies \citep{Seta2011}. These currently make up the bulk of the foreground lens galaxy population \citep[however, see][]{Narasimha2008}. In this regard, we plan to develop our technique further in a follow-up publication where the effect of a random magnetic field component is included in the model. We can also take into account the physical processes internal to the background source. In principle, one could fit any model to the reconstructed source to infer its intrinsic polarised and magnetic-field properties, similarly to what is commonly done in the analysis of non-lensed objects. In practice, however, this approach is problematic for strong gravitational lensing observations because the pixels are correlated on the source plane, the noise is not well characterised, and the effective resolution changes across the source due to a differential lensing magnification. These limitations could be overcome by introducing physically-motivated models into the prior of the pixellated source following the formalism presented by \citet{Rizzo2018}. 

In this paper, we have assumed that the polarised flux density, polarisation fraction and polarisation angle of the background source are constant in time. This is certainly the case for lensed star-forming galaxies \citep{Geach2023}, but for lensed AGN, any variability of the background source within the observing period and on scales smaller than the gravitational lensing time-delay \citep{Biggs2023} can introduce a bias in the inferred properties of the magnetised plasma in the lens as well as those of the source. We plan to investigate whether source variability can be included in the forward model. At the same time, from the expected large numbers of lensed sources to be found with next-generation instruments, it will be possible to select samples of objects that are not strongly variable in their polarised emission. 

We note that a dark-matter field made of axion-like particles also induces a differential rotation of the lensed-image polarisation angles \citep{Basu2021}. However, this effect is frequency-independent and, therefore, separable from that caused by the lens with observations taken with an appropriate frequency coverage.

For simplicity, our simulated data sets have a single Gaussian source that is polarised in its entirety for two different source sizes. The extent of the polarised emission has no effect on our ability to recover the lens model parameters since these are constrained using the Stokes $\vec{I}$ channel; this will always have a larger or equal extent and a larger signal-to-noise ratio when compared to the Stokes $\vec{Q}$ and $\vec{U}$ emission. We do find that our ability to recover the magneto-ionic parameters is affected by the extent of the background source (and its signal-to-noise ratio) since a larger source probes a larger part of the foreground lens. We have also limited our focus to one specific, though realistic, model for the lens galaxy (both in terms of the mass distribution and magneto-ionic properties) and data configuration (in terms of frequency coverage and angular resolution). We will provide an investigation of more generalised scenarios in the future. We will also investigate whether the data contains enough information to rank different choices for the magnetic field and electron density models. Finally, we have neglected the contribution of the lens environment. As many lens galaxies (though not all) reside within a galaxy cluster or group, one should expect a non-negligible contribution to the Faraday rotation by the intergalactic medium \citep[][]{Greenfield1985}. In the future, the large samples of gravitational lens systems provided by LOFAR2.0 and SKA-MID could allow one to study the Faraday rotation effect as a function of the lens properties and environment. We intend to test this using simulations of SKA-like observations in the future.

\section{Conclusions}
\label{sec:summary}

Existing and upcoming observing facilities such as LOFAR2.0 and the SKA are characterised by a large sensitivity, field of view and bandwidth, and a large number density of polarised sources. They are expected, therefore, to push the boundaries of our knowledge on magnetic fields in a large variety of astrophysical systems, from the Milky Way and high-redshift radio galaxies to fast radio bursts and the cosmic web \citep{Beck2007,Beck2011,Heald2020}. However, the polarised synchrotron emission from galaxies at high-redshift ($z\sim$2 and above) will be too faint to be easily detected even with these telescopes, and measuring their magnetic field structure directly will be challenging. Observations of polarised strongly-lensed galaxies will allow us to study magnetic fields at cosmological distances that cannot otherwise be probed. At the same time, the effect of Faraday rotation within the lens will also provide a measure of the magneto-ionic properties of the deflector. Similarly to the traditional method of Faraday rotation measure towards background polarised sources, strong gravitational lensing provides a partial view of the magnetic field in the intervening galaxy. However, compared to the original approach, it has the advantage of being a differential measurement and, hence, potentially more robust on an object-by-object basis. Moreover, as most of the lens galaxies will be massive ellipticals, strong gravitational lensing provides us with an avenue to study a population of objects for which magnetic fields are still poorly understood and can potentially provide observational insights into the fluctuation dynamo action \citep{Seta2011}. Uniquely, our method can probe the magneto-ionic properties of two objects at two different redshifts (the lens and the source) from a single observation. 

In this paper, we introduced the first fully self-consistent framework to simultaneously constrain the mass density distribution, the magneto-ionic properties of strong gravitational lens galaxies, and the polarised surface brightness distribution of the background sources in the native visibility plane. We tested this modelling technique using simulated data sets with a realistic lens and source at GHz frequencies, different levels of source polarisation fraction and different lensing configurations. We found that we can recover the rotation measure from the lens galaxy with an accuracy that is dependent on the lensing configuration, the underlying magnetic field model and the data signal-to-noise ratio.  We also obtain a reliable reconstruction of the polarised source properties with better performance for higher signal-to-noise ratios of the data and a larger fraction of the source within the diamond caustic. In the future, we will further develop our method to address some of its current limitations and use existing and upcoming data to investigate the nature of magnetic fields at cosmological distances.

\section*{Acknowledgements}
S.N., S.V., and D.M.P. have received funding from the European Research Council (ERC) under the European Union's Horizon 2020 research and innovation programme (grant agreement No 758853). S.V. thanks the Max Planck Society for support through a Max Planck Lise Meitner Group. JPM acknowledges support from the Netherlands Organization for Scientific Research (NWO) (Project No. 629.001.023) and the Chinese Academy of Sciences (CAS) (Project No. 114A11KYSB20170054). This work is based on the research supported in part by the National Research Foundation of South Africa (Grant Number: 128943).

\section*{DATA AVAILABILITY}
The data used in this paper are available from the corresponding author on request.




\bibliographystyle{mnras}
\bibliography{ms}



\appendix
\section{Appendix}
\subsection{Reconstructions for all data sets}
\label{sect:plots}
\begin{figure*}
\centering
\includegraphics[scale=0.6]{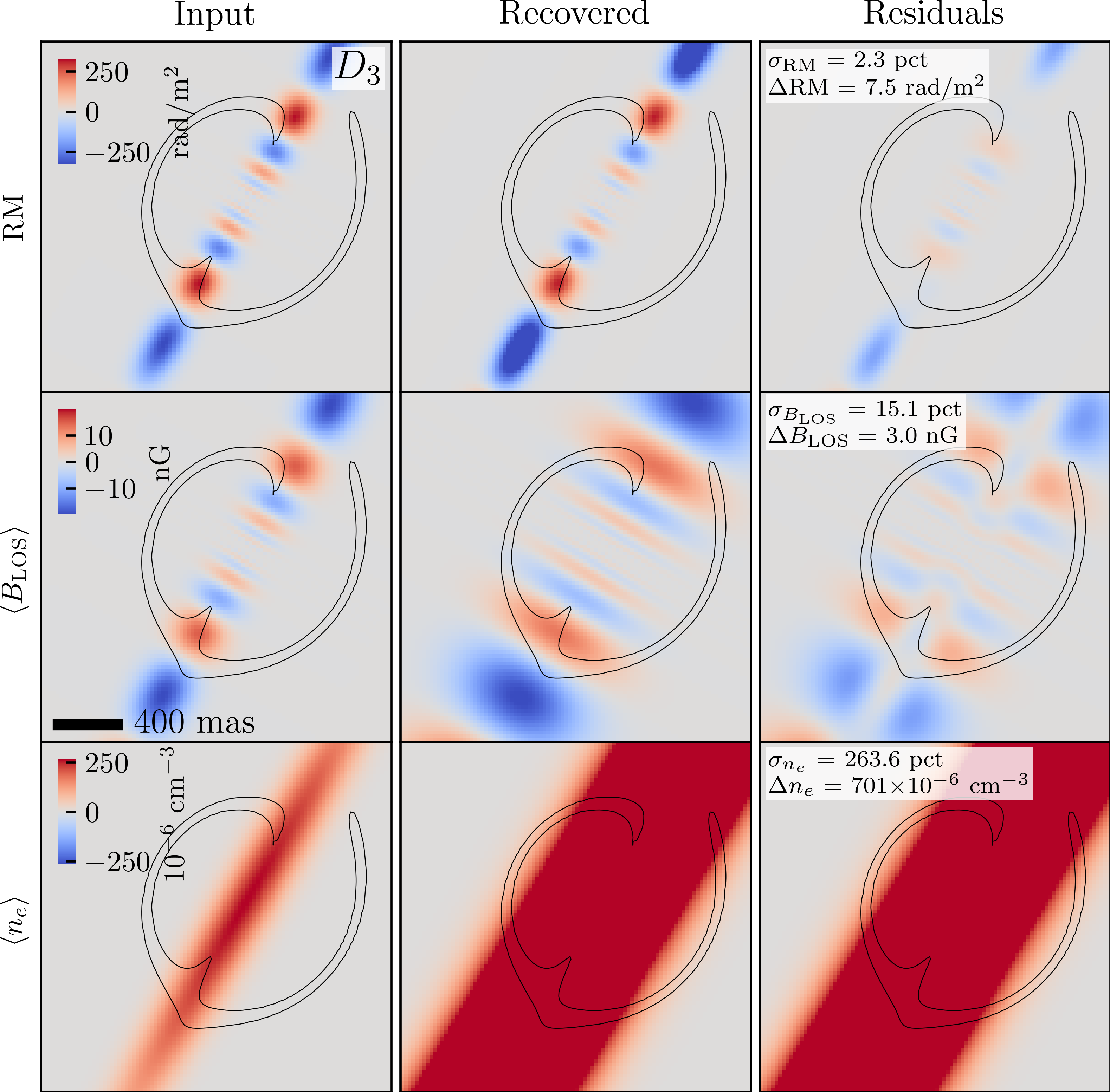}
 \includegraphics[scale=0.6]{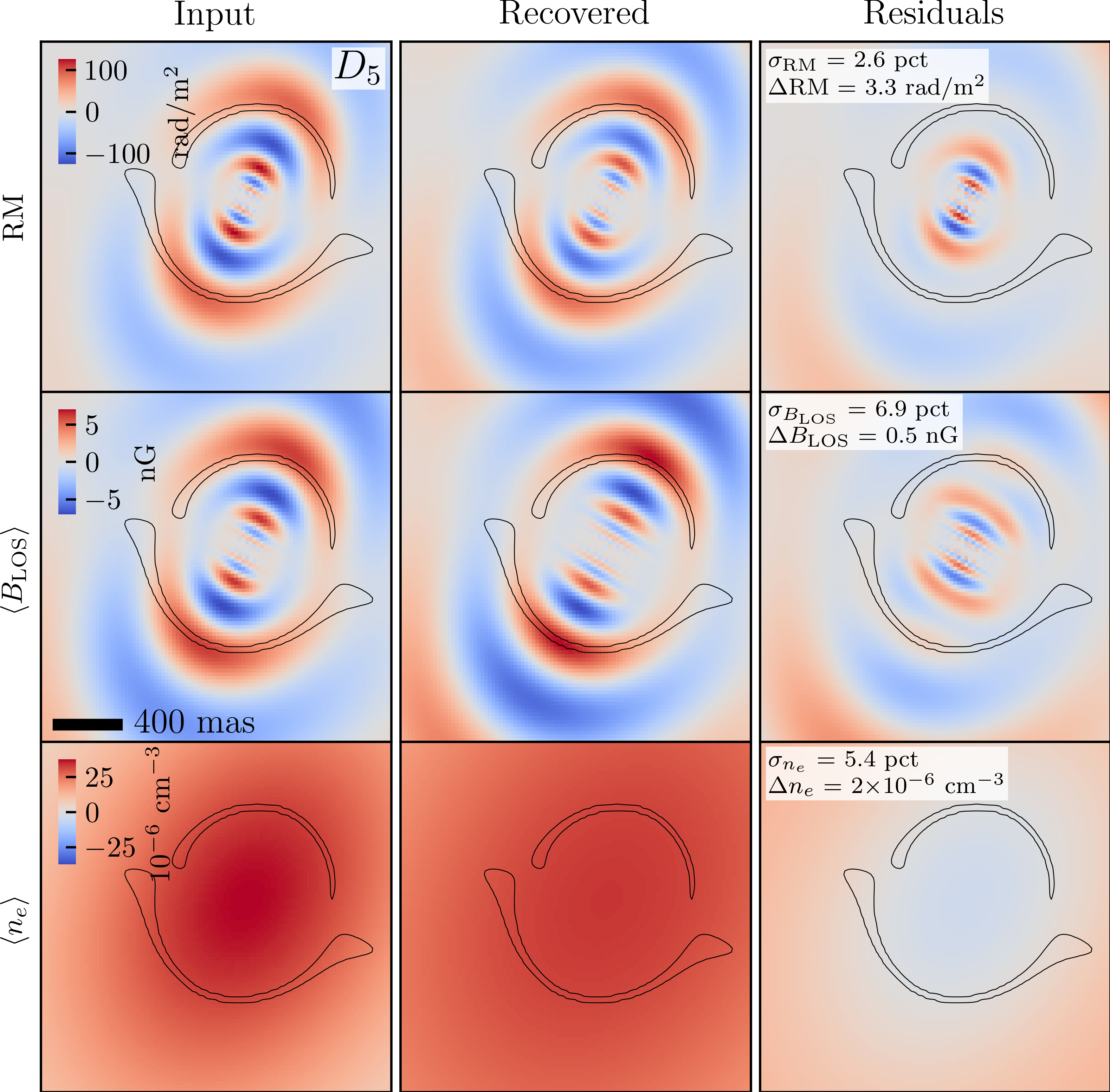}
\caption{Same as Figure \ref{fig:b_ne_rm_comparison_d1d8} for D$_{\rm 3}$ and D$_{\rm 5}$.}
\label{fig:b_ne_rm_comparison_app_1}
\end{figure*}

\begin{figure*}
\centering
\includegraphics[scale=0.6]{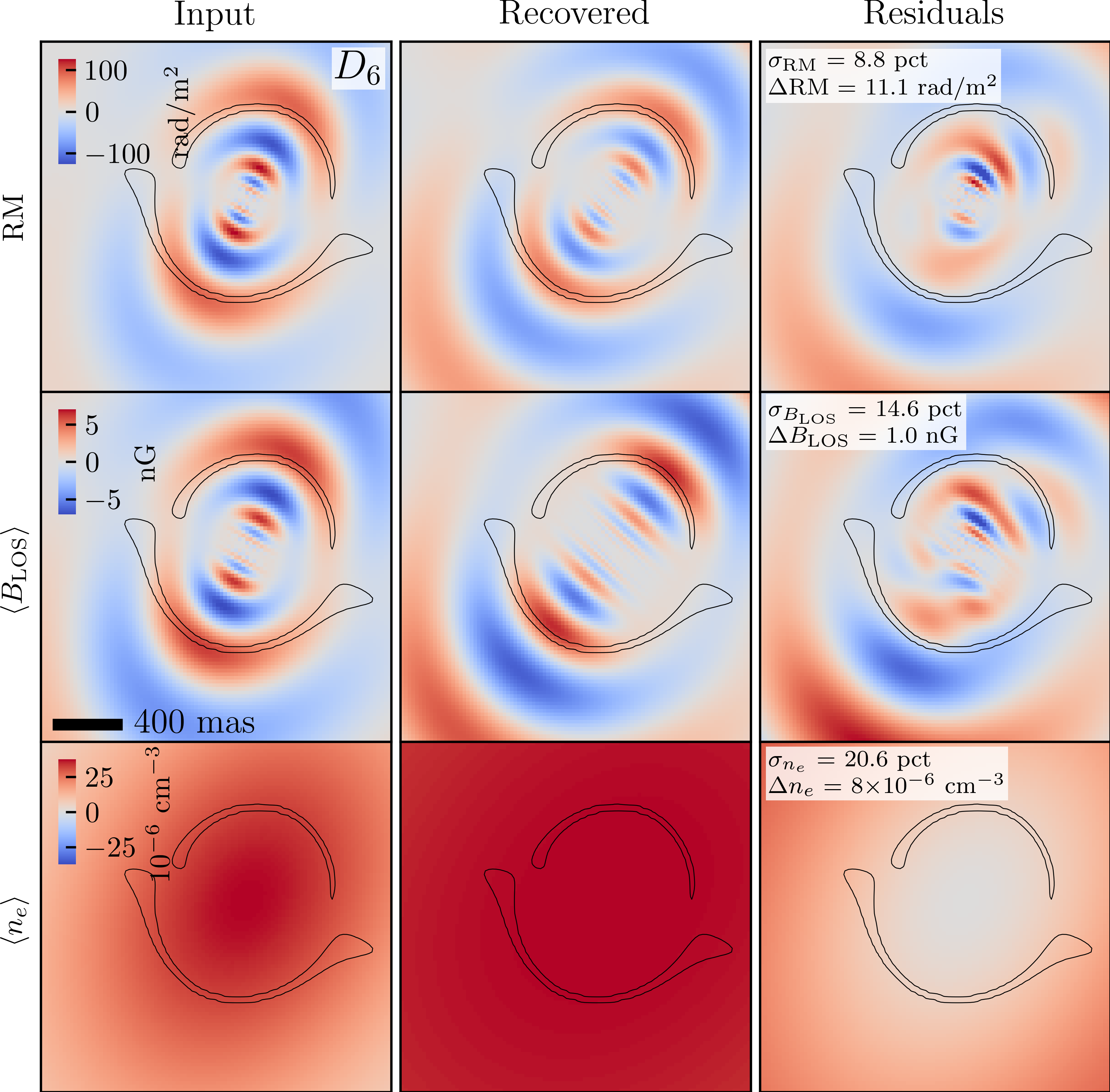}
\includegraphics[scale=0.6]{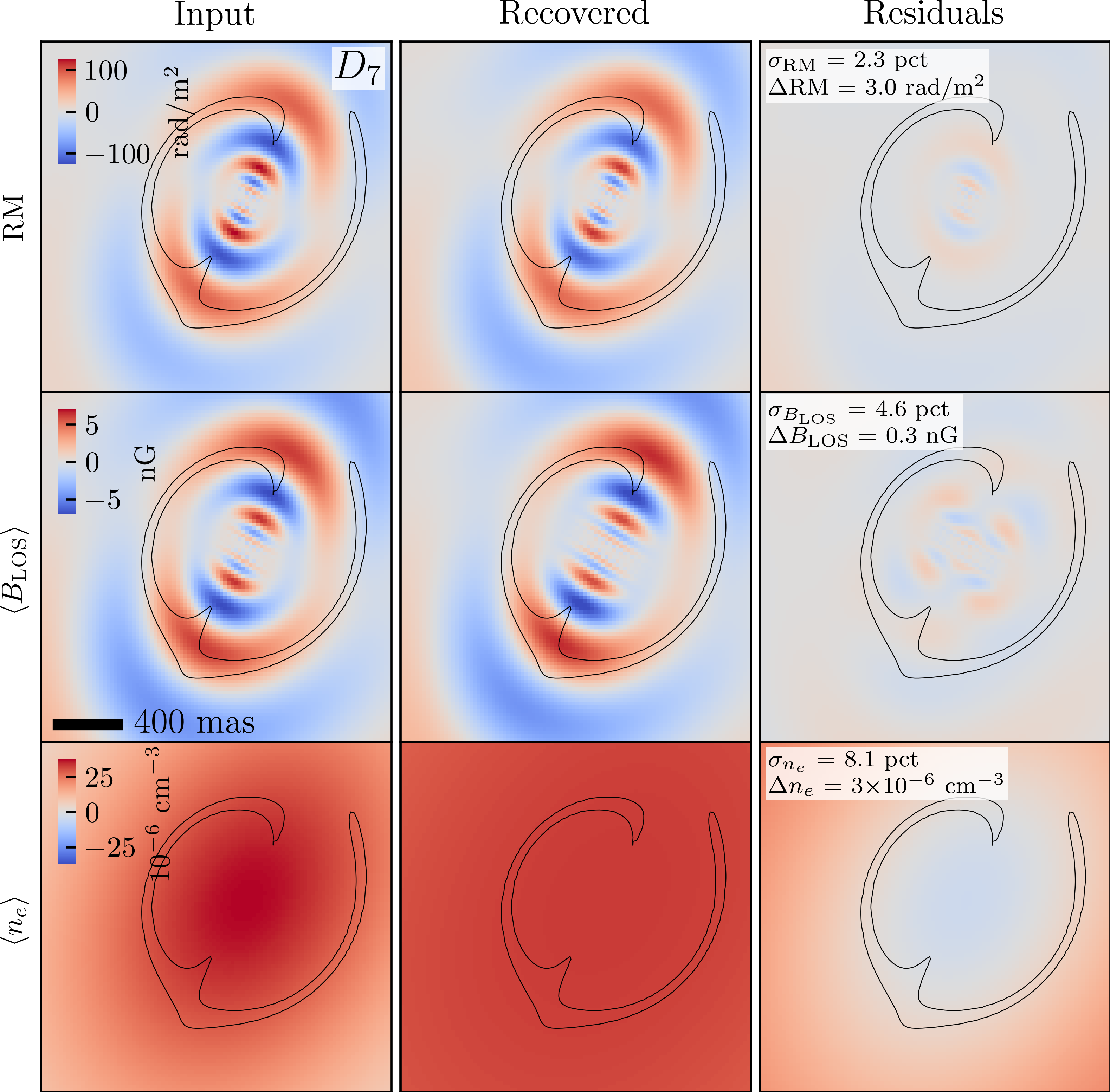}~~
\caption{Same as Figure \ref{fig:b_ne_rm_comparison_d1d8} for D$_{\rm 6}$ and D$_{\rm 7}$.}
\label{fig:b_ne_rm_comparison_app_2}
\end{figure*}

\begin{figure*}
\centering
 \includegraphics[angle= 90,width=0.9\textwidth]{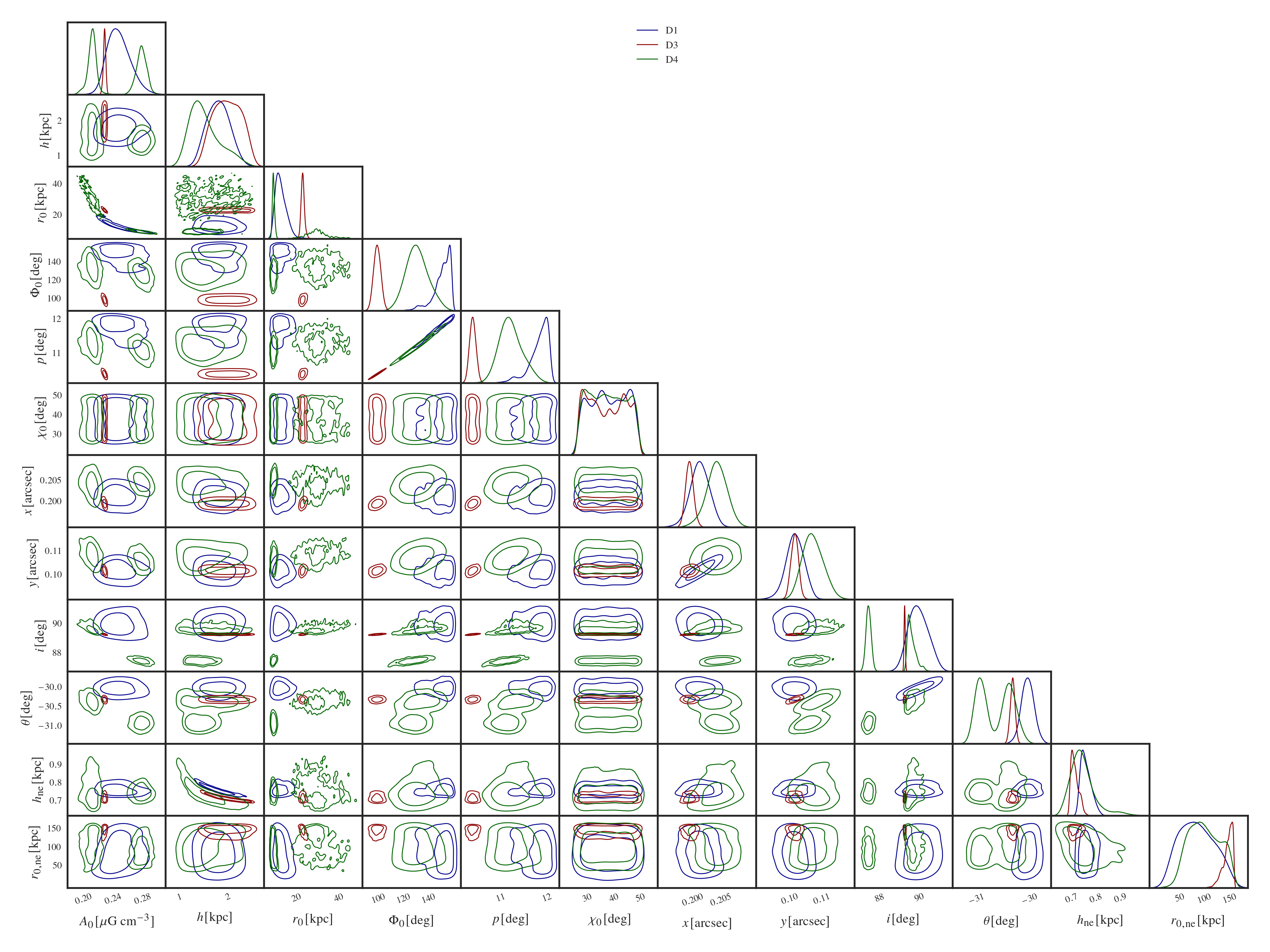}
\caption{Simulated data sets D$_{\rm 1}$, D$_{\rm 3}$ and D$_{\rm 4}$. The posterior distribution for all the parameters of the lens magnetic field and electron density.}
\label{fig:mn_90deg}
\end{figure*}

\begin{figure*}
\centering
 \includegraphics[angle= 90,width=0.9\textwidth]{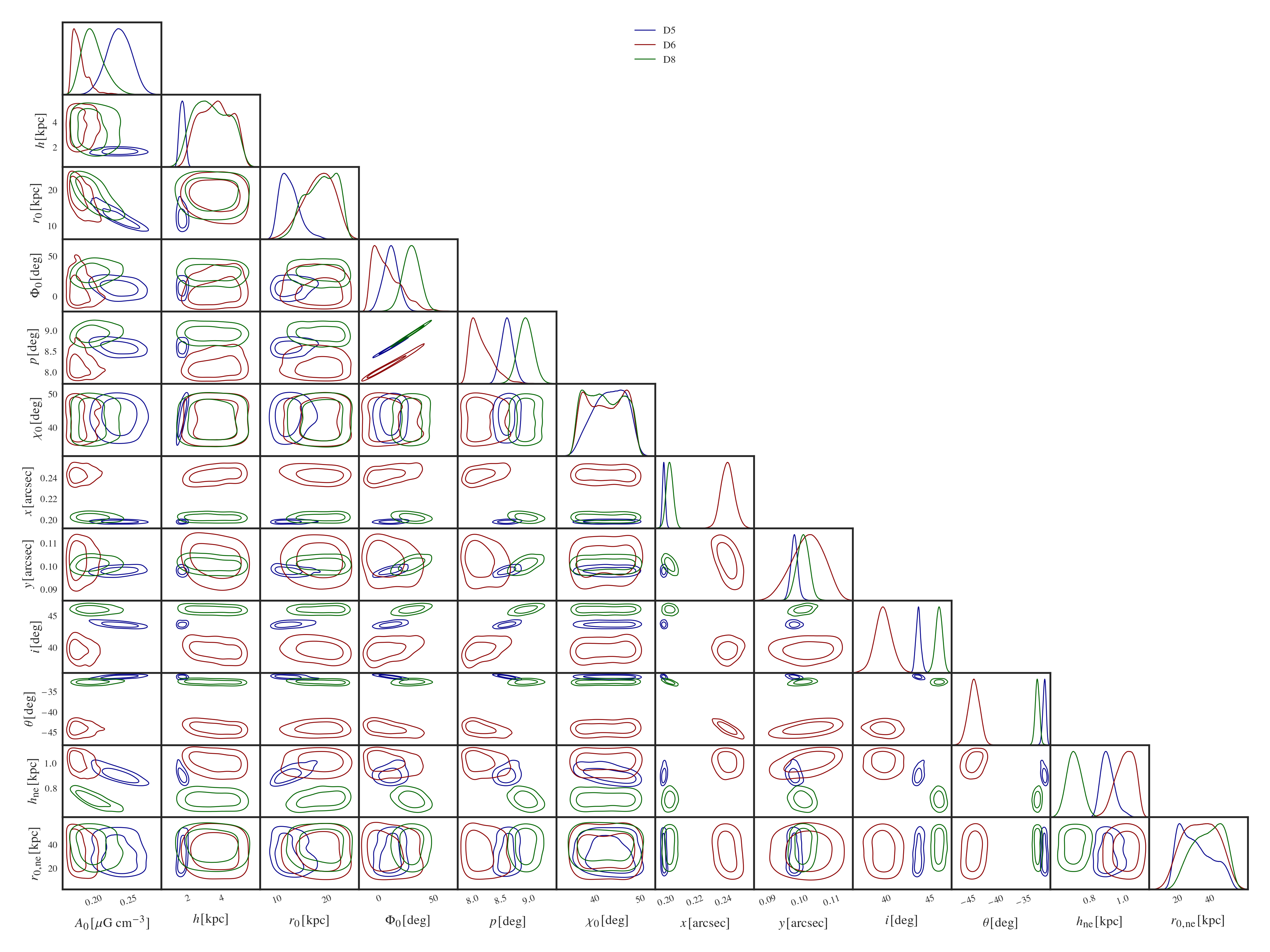}
\caption{Simulated data sets D$_{\rm 5}$, D$_{\rm 6}$ and D$_{\rm 8}$. The posterior distribution for all the parameters of the lens magnetic field and electron density.}
\label{fig:mn_45deg}
\end{figure*}

\begin{figure*}
\centering
\includegraphics[width=0.9\textwidth]{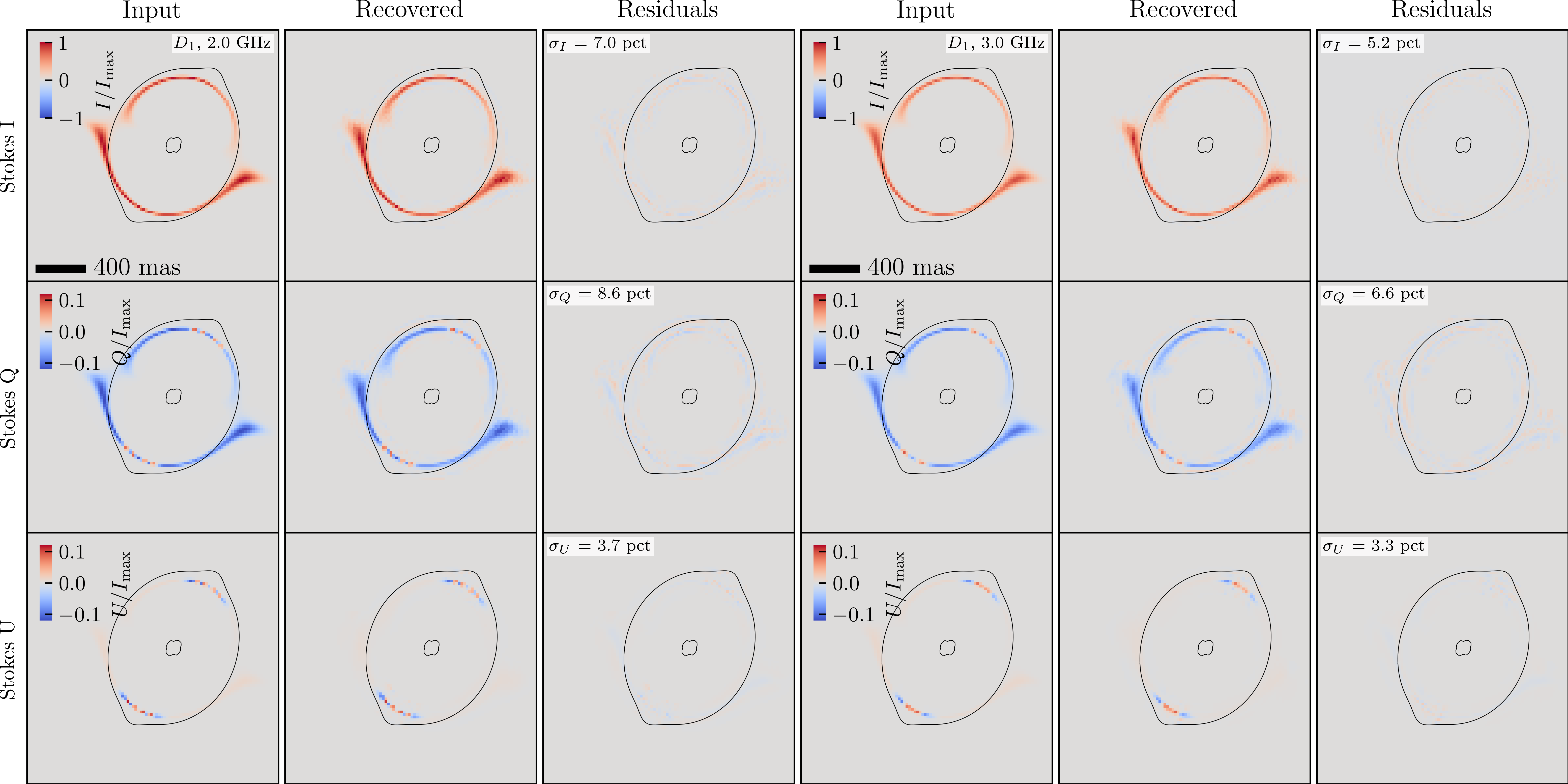}\\~\\
\includegraphics[width=0.9\textwidth]{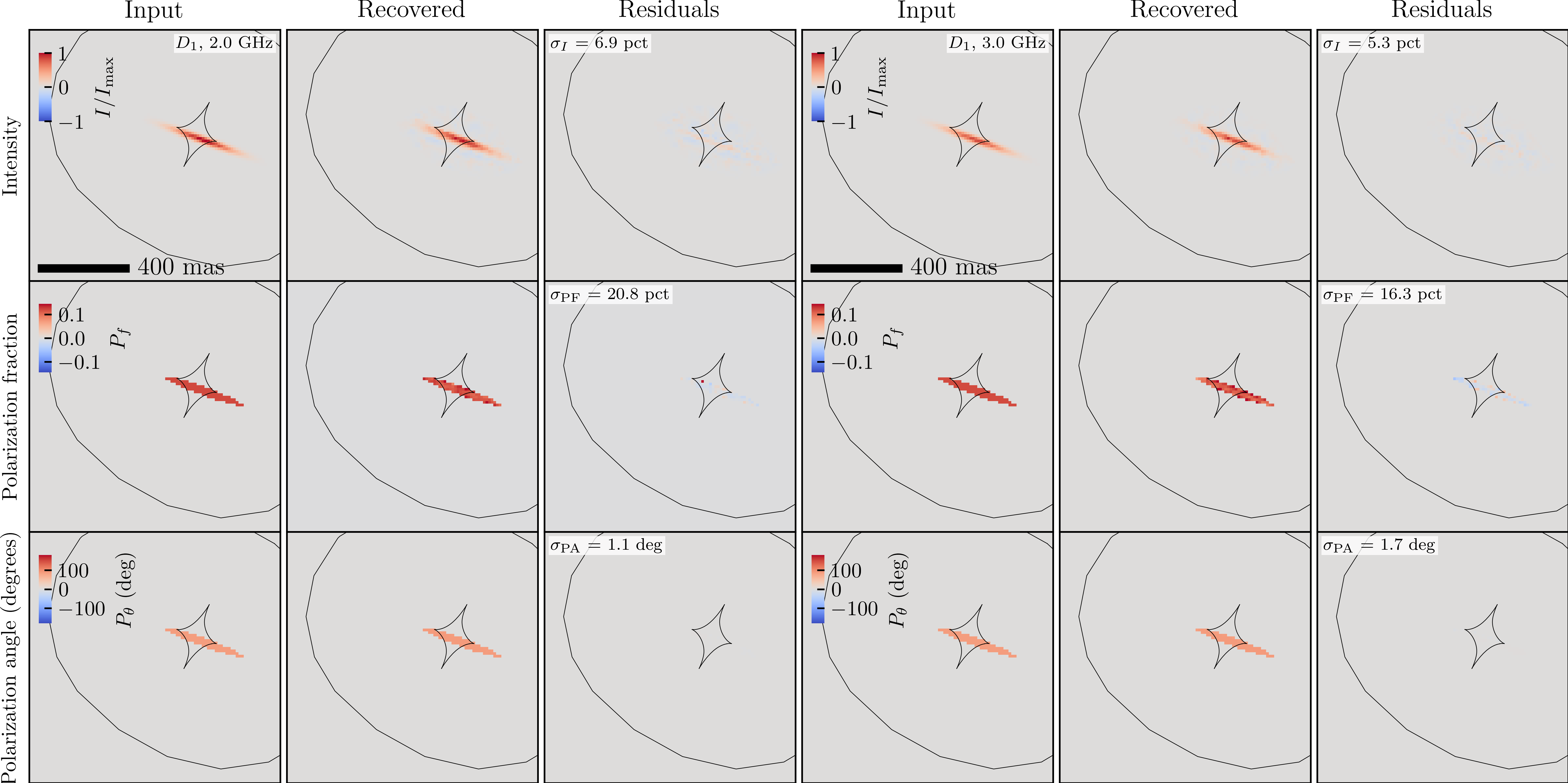}\\~\\
\caption{Data set $D_{\rm 1}$. Top: comparison of image-plane results with the ground-truth input at two representative frequency channels, as an example. The rows contain Stokes I (top), the polarisation fraction (middle), and the polarisation angle (in deg; bottom). The rows contain Stokes I, Q, and U from top to bottom. Bottom: Comparison of source-plane results with the ground-truth input. The rows contain Stokes I (top), the polarisation fraction (middle), and the polarisation angle (in deg; bottom).}
\label{fig:SLs_comparison_app_1}
\end{figure*}

\begin{figure*}
\centering
\includegraphics[width=0.9\textwidth]{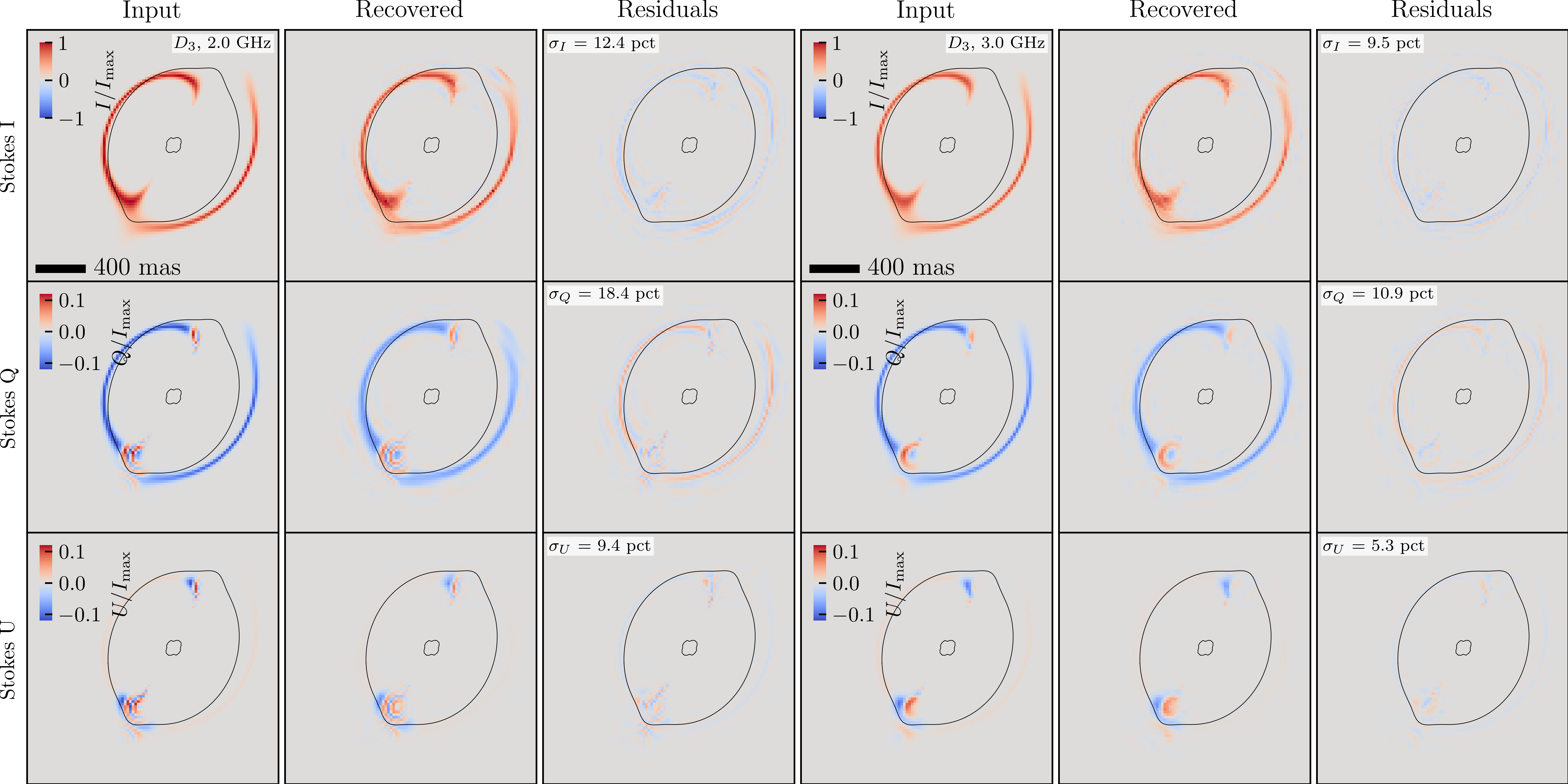}\\~\\
\includegraphics[width=0.9\textwidth]{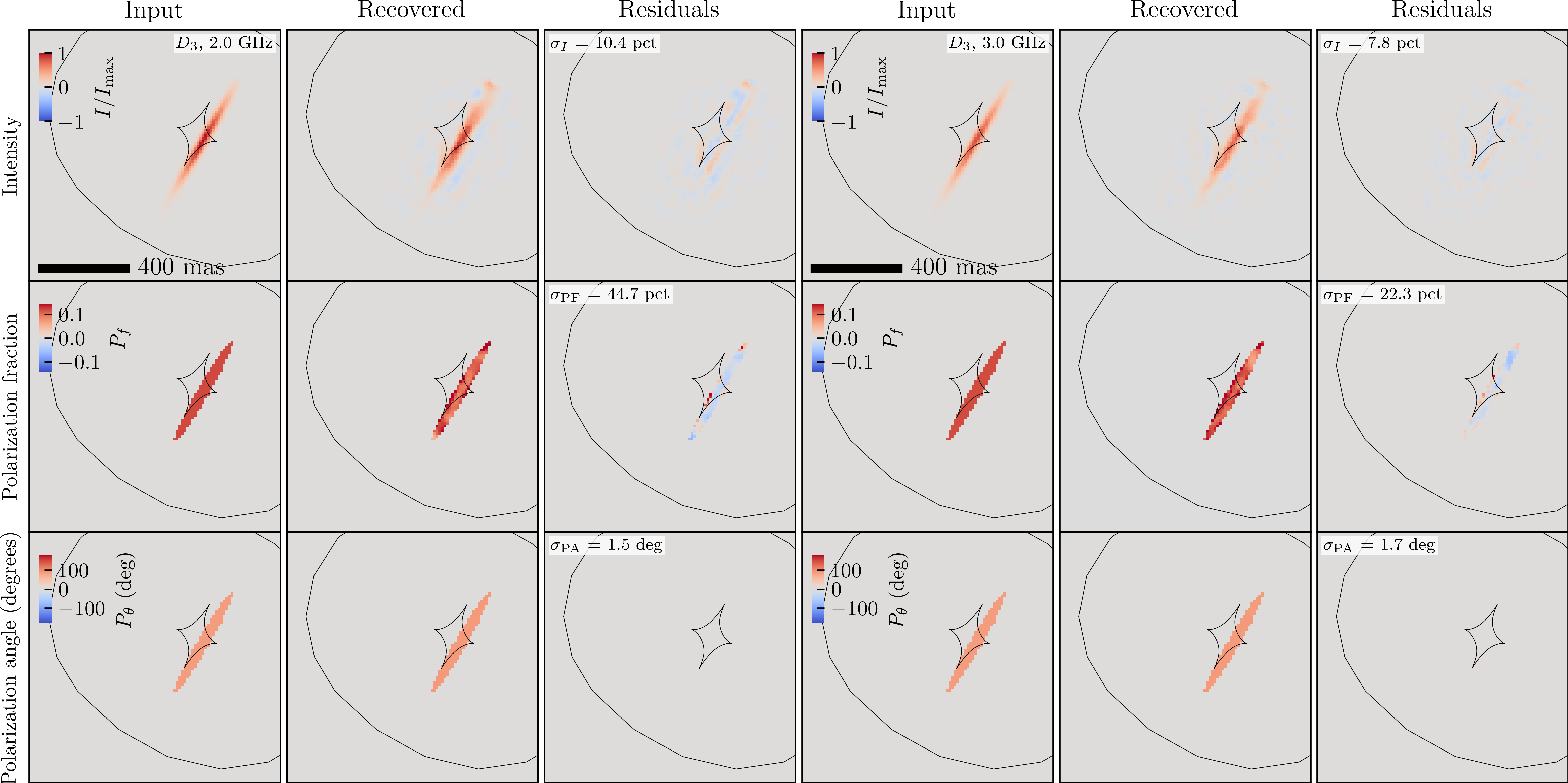}\\~\\
\caption{Same as Figure \ref{fig:SLs_comparison_app_1} for D$_{\rm 3}$.}
\label{fig:SLs_comparison_app_2}
\end{figure*}

\begin{figure*}
\centering
\includegraphics[width=0.9\textwidth]{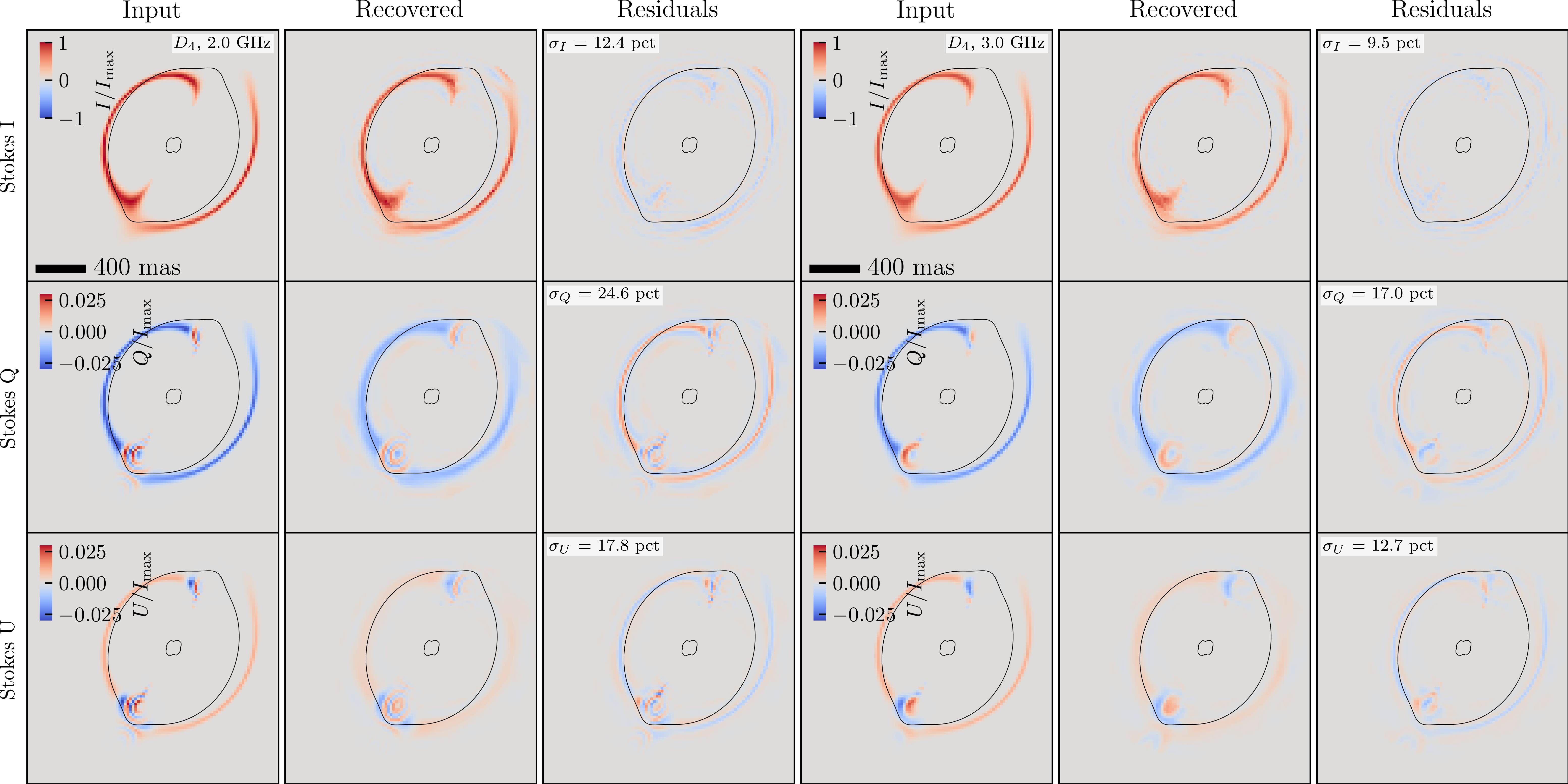}\\~\\
\includegraphics[width=0.9\textwidth]{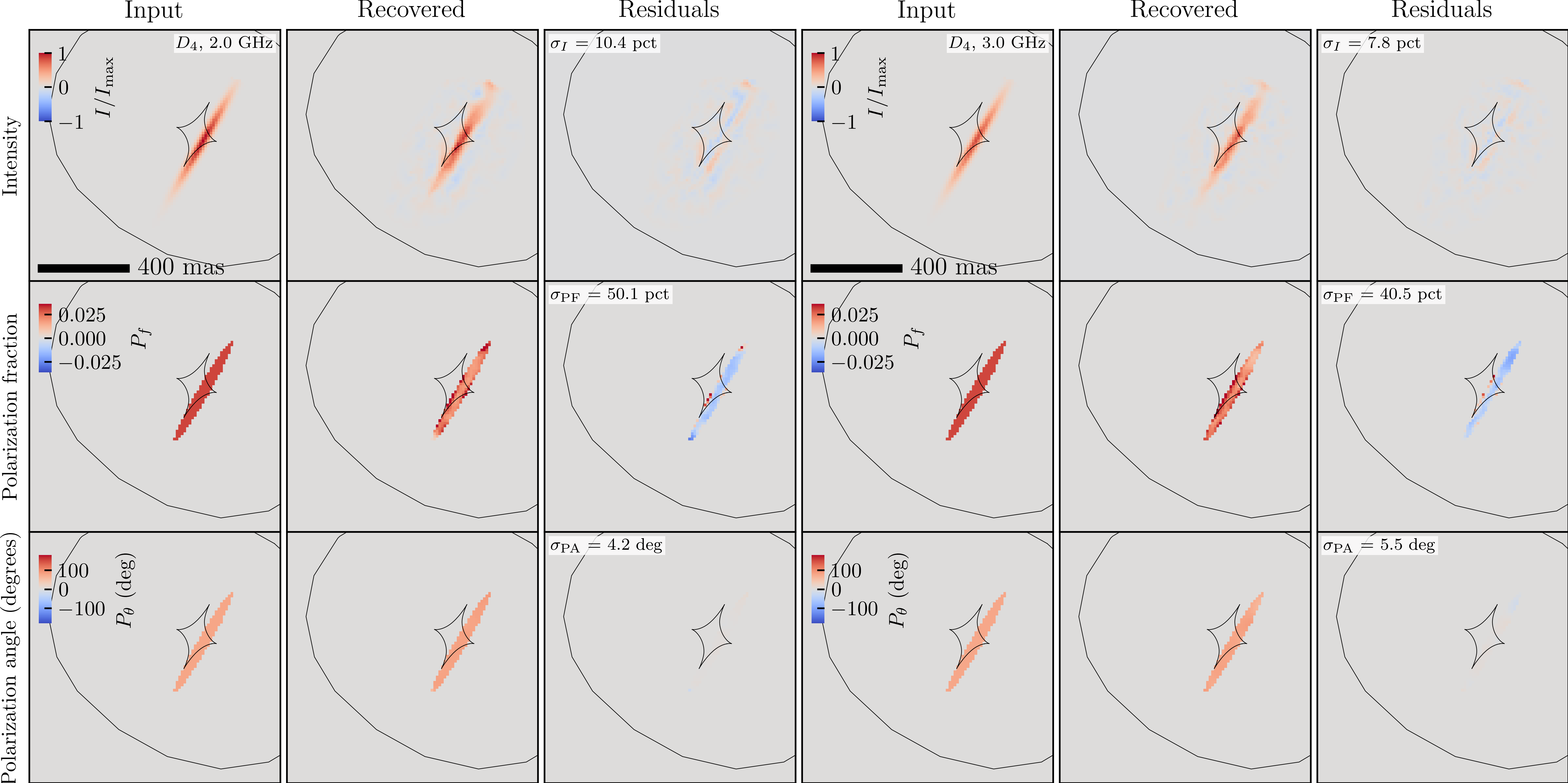}\\~\\
\caption{Same as Figure \ref{fig:SLs_comparison_app_1} for D$_{\rm 4}$.}
\label{fig:SLs_comparison_app_3}
\end{figure*}

\begin{figure*}
\centering
\includegraphics[width=0.9\textwidth]{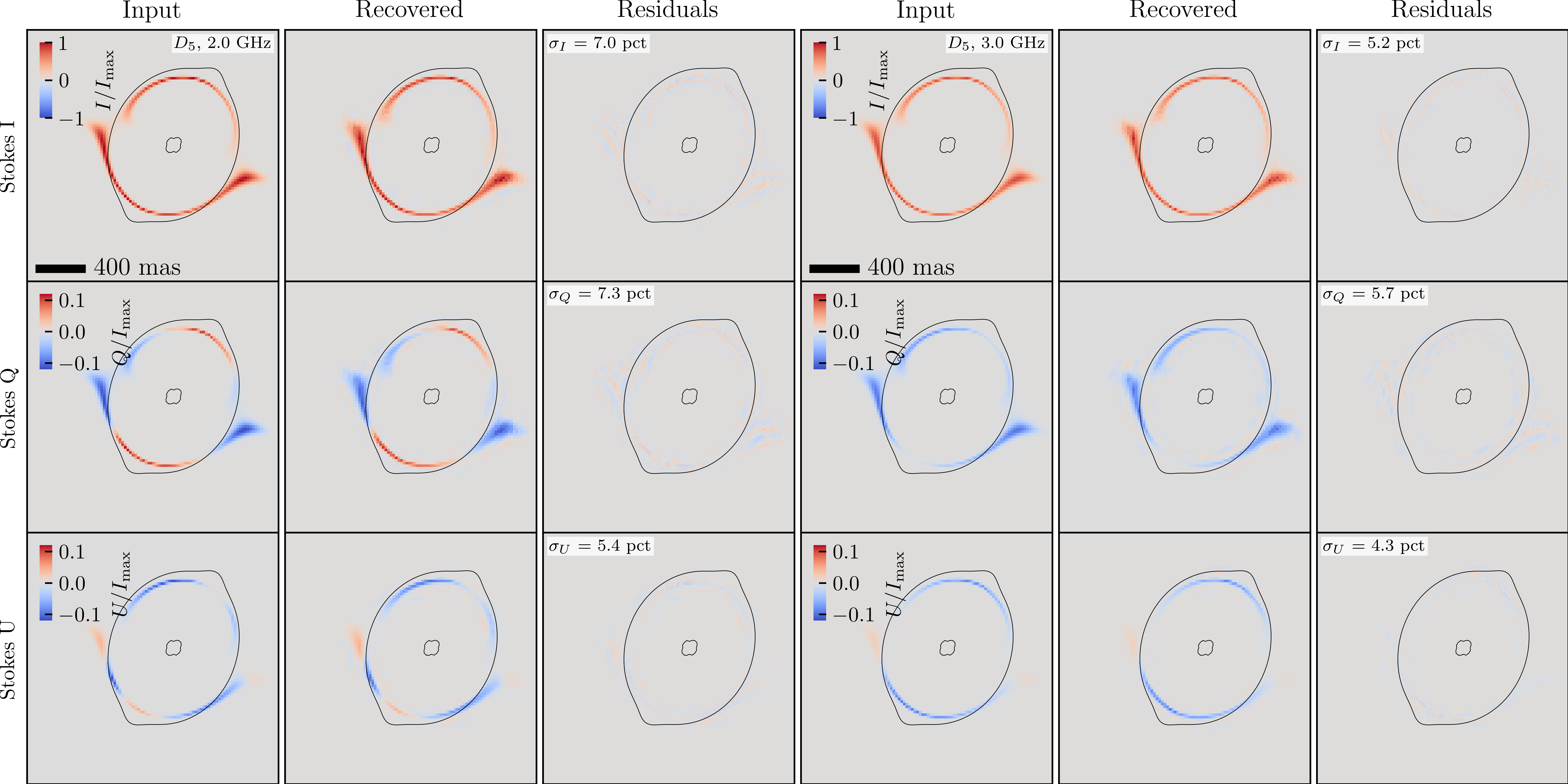}\\~\\
\includegraphics[width=0.9\textwidth]{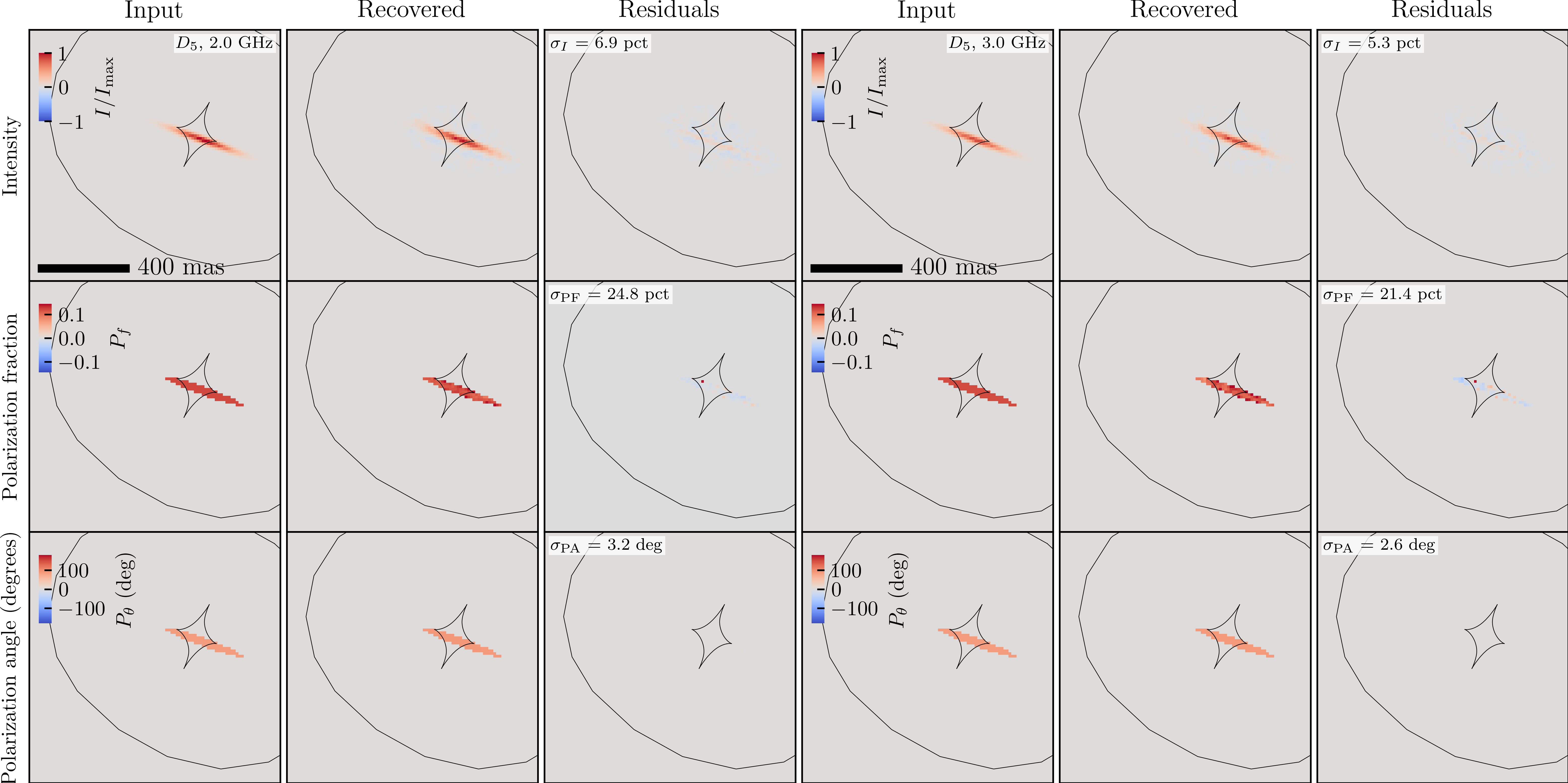}\\~\\
\caption{Same as Figure \ref{fig:SLs_comparison_app_1} for D$_{\rm 5}$.}
\label{fig:SLs_comparison_app_4}
\end{figure*}

\begin{figure*}
\centering
\includegraphics[width=0.9\textwidth]{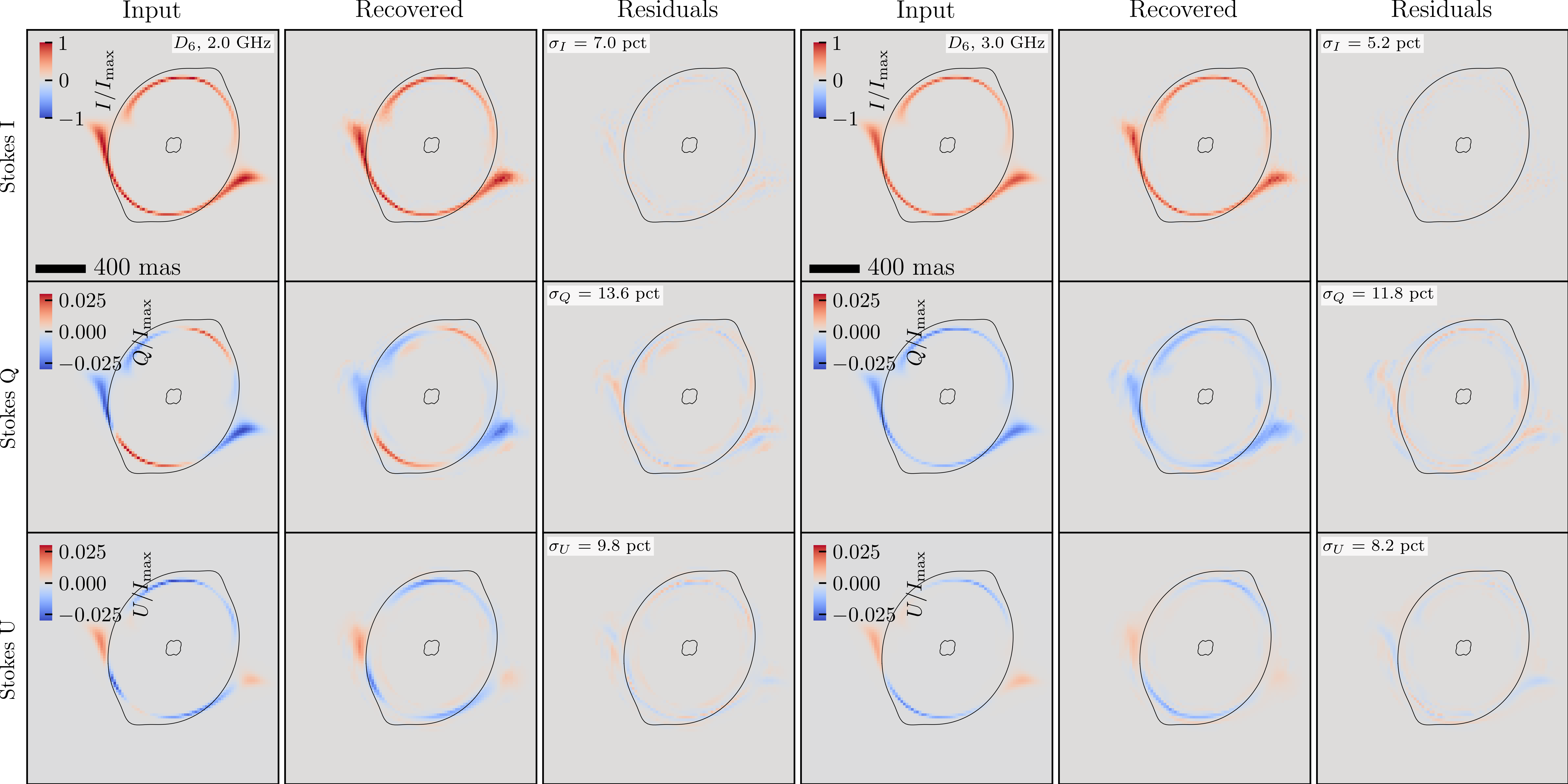}\\~\\
\includegraphics[width=0.9\textwidth]{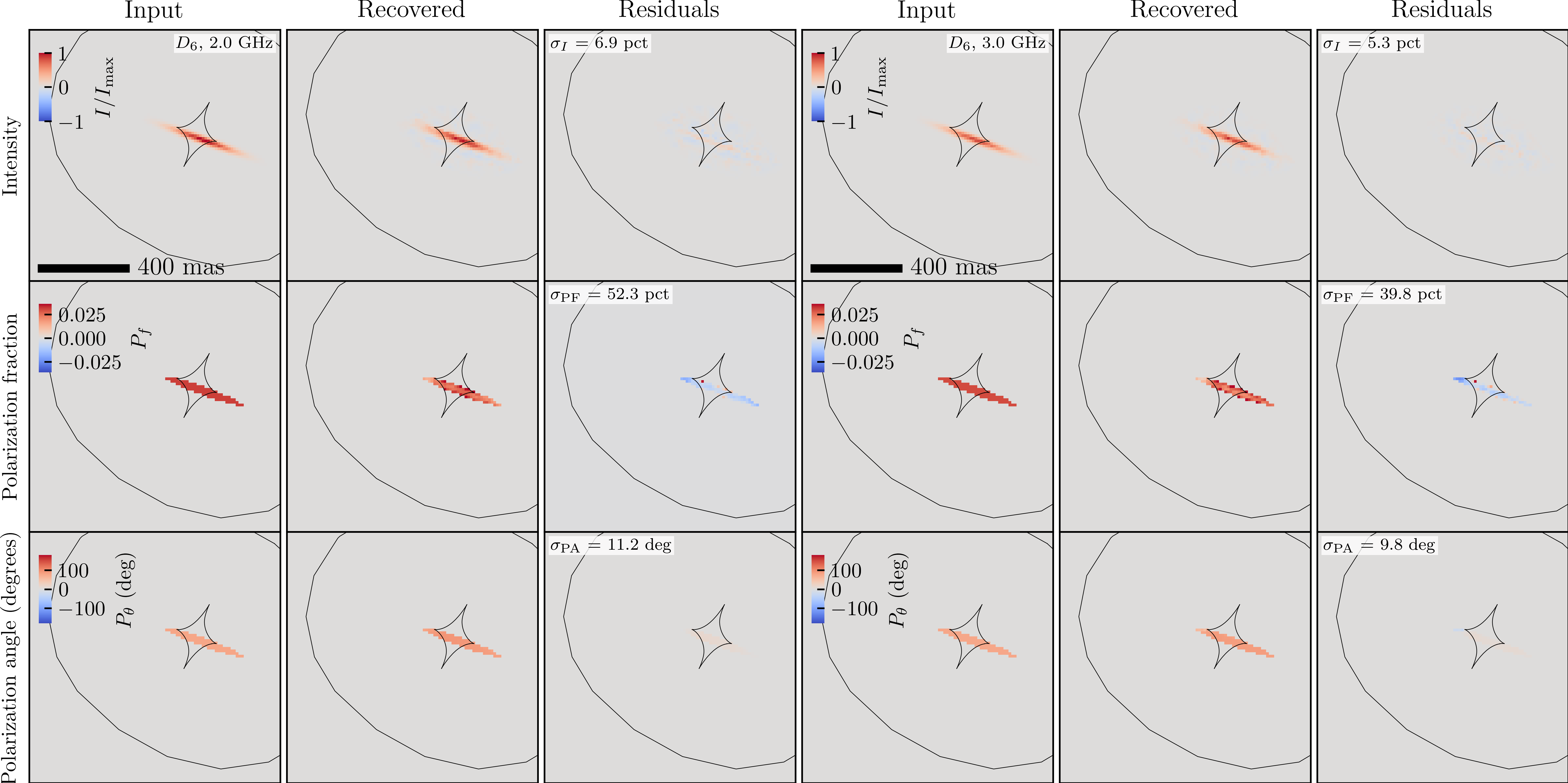}\\~\\
\caption{Same as Figure \ref{fig:SLs_comparison_app_1} for D$_{\rm 6}$.}
\label{fig:SLs_comparison_app_5}
\end{figure*}

\begin{figure*}
\centering
\includegraphics[width=0.9\textwidth]{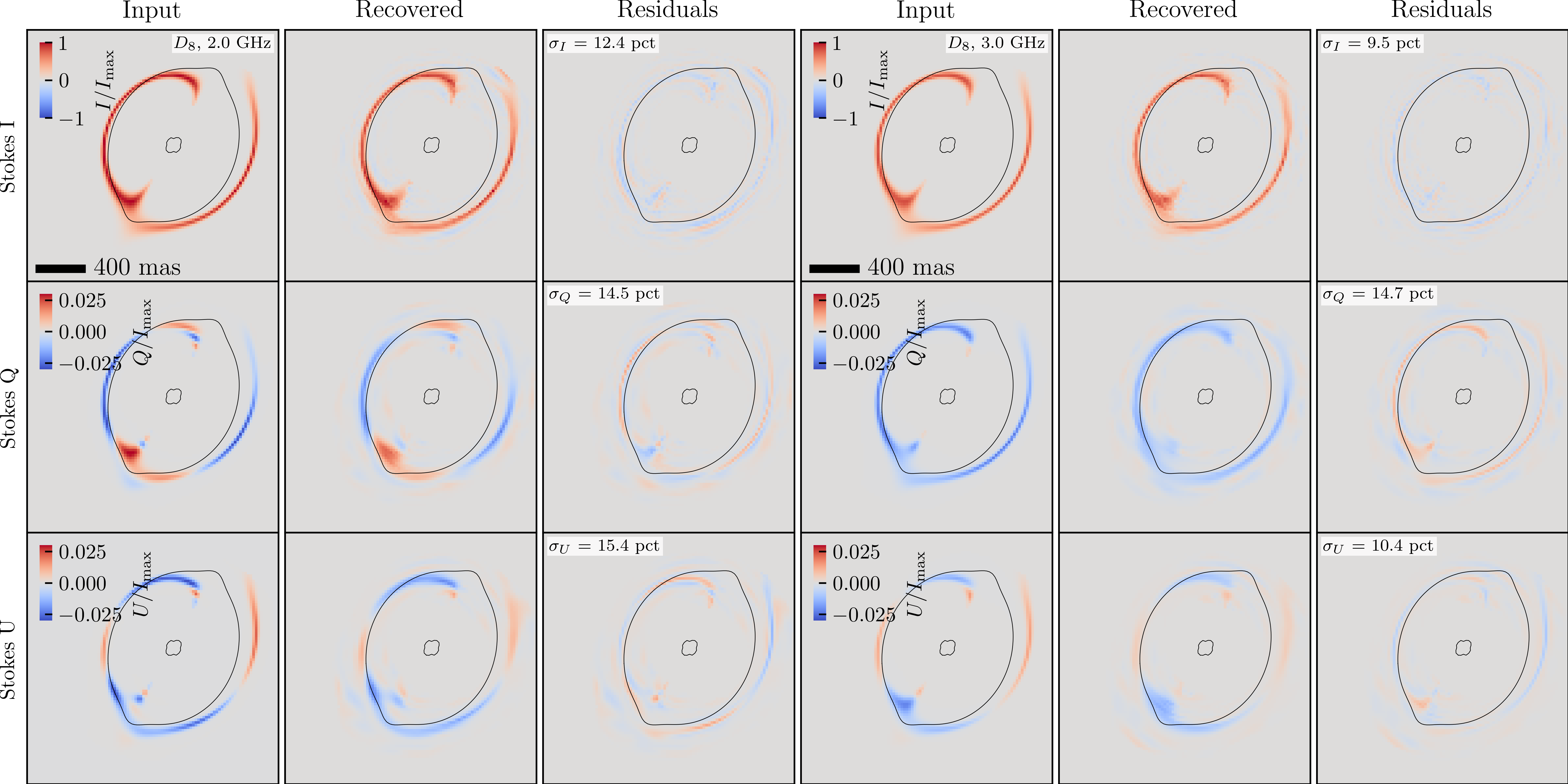}\\~\\
\includegraphics[width=0.9\textwidth]{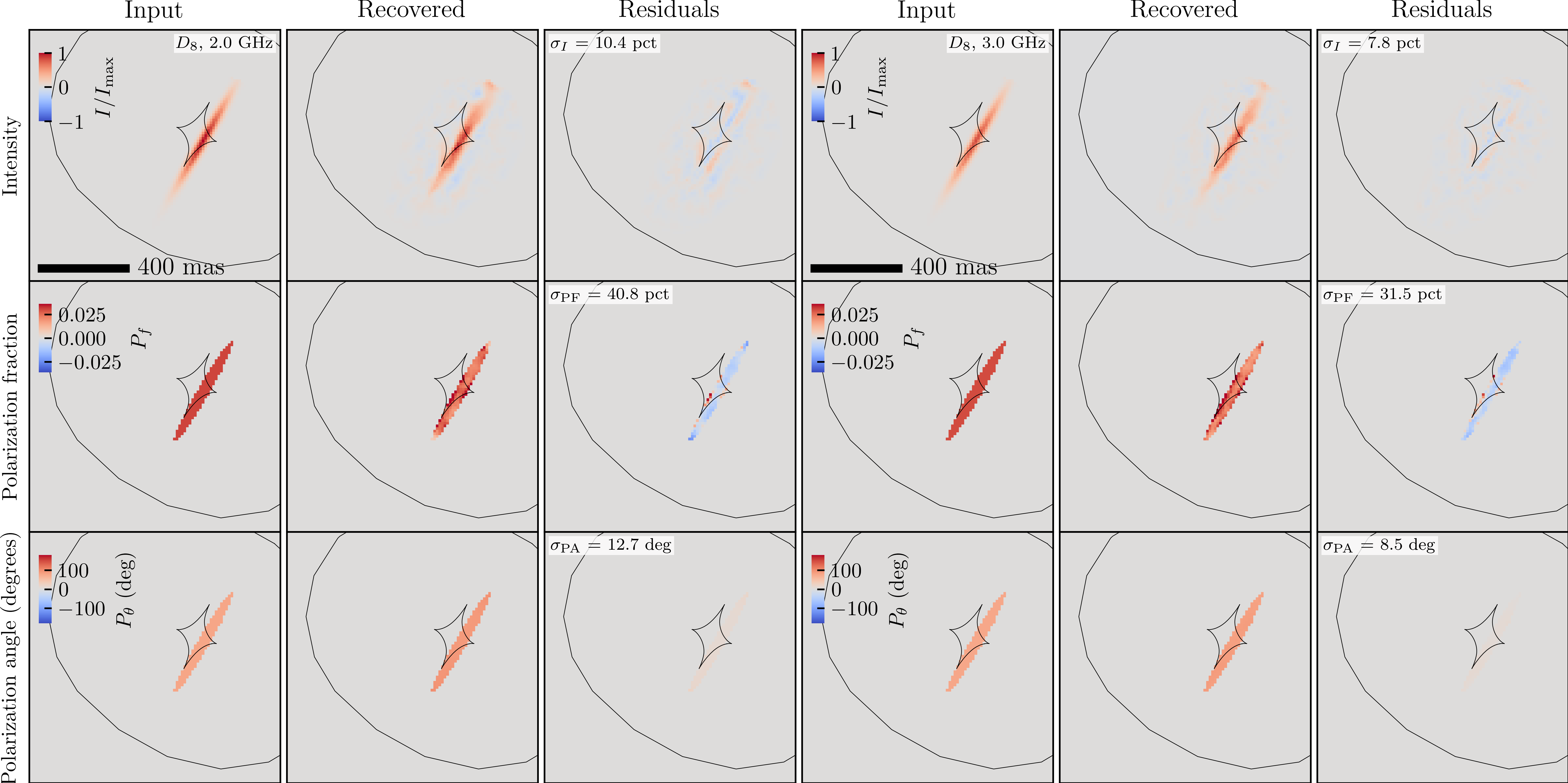}\\~\\
\caption{Same as Figure \ref{fig:SLs_comparison_app_1} for D$_{\rm 8}$.}
\label{fig:SLs_comparison_app_6}
\end{figure*}

This section displays the modelling results for all the data sets not shown in the main text. In particular, Figs. \ref{fig:b_ne_rm_comparison_app_1} and \ref{fig:b_ne_rm_comparison_app_2} display the inferred rotation measure, electron density and line-of-sight magnetic field compared to the input models. The Figs. \ref{fig:mn_90deg} and \ref{fig:mn_45deg} show the posterior distribution for the magnetic field and electron density parameters for the data sets with an inclination angle of 90 and 45 deg, respectively. The reconstructed and input \software{CLEAN}ed lensed images and sources are plotted in Figs. \ref{fig:SLs_comparison_app_1} to \ref{fig:SLs_comparison_app_6}.

\bsp	
\label{lastpage}
\end{document}